\documentclass[twocolumn]{aastex63}
\usepackage{ulem,color}
\usepackage{amsmath,amssymb,bm}
\usepackage{etoolbox}
\usepackage{booktabs}
\usepackage{natbib}

\def \del2z {\partial^{2}_{z}}

\def \A {{\bm A}}

\def \f {{\bm f}}

\def \B {{\bm B}}

\def \k {{\bm k}}
\def \x {{\bm x}}

\def \t  {\theta}
\def \p  {\phi}

{}

\def \B {\bm B}

\def \A {\bm A}

 \def \a {{\alpha_0}}

\def\drawing #1 #2 #3 {
\begin{center}
\setlength{\unitlength}{1mm}
\begin{picture}(#1,#2)(0,0)
\put(0,0){\framebox(#1,#2){#3}}
\end{picture}
\end{center} }

\usepackage[normalem]{ulem}

\newcommand{\be}{\begin{equation}}
\newcommand{\ee}{\end{equation}}
\def\bear#1\ear{\begin{align}#1\end{align}}

\newcommand{\e}{\mathrm{e}}

\renewcommand{\mathbf}[1]{\mbox{\boldmath $#1$}}

\newcommand{\Real}{\mathrm{Re}\,}
\newcommand{\Imag}{\mathrm{Im}\,}
\newcommand{\X}{\mathcal{X}}
\newcommand{\Y}{\mathcal{Y}}
\newcommand{\Z}{\mathcal{Z}}
\newcommand{\T}{\Theta}

\def\vecsign{\mathchar"017E}
\def\dvecsign{\smash{\stackon[-1.95pt]{\vecsign}{\rotatebox{180}{$\vecsign$}}}}
\def\dvec#1{\def\useanchorwidth{T}\stackon[-4.2pt]{#1}{\,\dvecsign}}
\usepackage{stackengine}
\stackMath
\usepackage{graphicx}
\usepackage[T1,T2A]{fontenc} 
\DeclareTextSymbolDefault{\ae}{T1}

\received{\today}
\shorttitle{Angular Momentum Transport Model}
\shortauthors{Tripathi, Barker, Fraser, Terry, \& Zweibel}
\graphicspath{{./}{figures/}}

\begin{document}
\title[Angular Momentum Transport Model]{Predicting the Slowing of Stellar Differential Rotation by Instability-Driven Turbulence}
\author[0000-0002-4723-2170]{B.~Tripathi}
\affiliation{Department of Physics, University of Wisconsin--Madison, Madison, Wisconsin 53706, USA}
\email{btripathi@wisc.edu}

\author[0000-0003-4397-7332]{A.J.~Barker}
\affiliation{Department of Applied Mathematics, School of Mathematics, University of Leeds, Leeds LS2 9JT, UK}
\email{A.J.Barker@leeds.ac.uk}

\author[0000-0003-4323-2082]{A.E.~Fraser}
\affiliation{Department of Applied Mathematics, University of Colorado, Boulder, Colorado 80309, USA}
\affiliation{Department of Astrophysical and Planetary Sciences, University of Colorado, Boulder, Colorado 80309, USA}
\affiliation{Laboratory for Atmospheric and Space Physics, University of Colorado, Boulder, Colorado 80309, USA}

\author[0000-0002-4981-9637]{P.W.~Terry}
\affiliation{Department of Physics, University of Wisconsin--Madison, Madison, Wisconsin 53706, USA}

\author[0000-0003-4821-713X]{E.G.~Zweibel}
\affiliation{Department of Physics, University of Wisconsin--Madison, Madison, Wisconsin 53706, USA}
\affiliation{Department of Astronomy, University of Wisconsin--Madison, Madison, Wisconsin 53706, USA}

\begin{abstract}
Differentially rotating stars and planets transport angular momentum internally due to turbulence at rates that have long been a challenge to predict reliably. We develop a self-consistent saturation theory, using a statistical closure approximation, for hydrodynamic turbulence driven by the axisymmetric Goldreich--Schubert--Fricke (GSF) instability at the stellar equator with radial differential rotation. 
This instability arises when fast thermal diffusion eliminates the stabilizing effects of buoyancy forces in a system where a stabilizing entropy gradient dominates over the destabilizing angular momentum gradient.
Our turbulence closure invokes a dominant three-wave coupling between pairs of linearly unstable eigenmodes and a near-zero frequency, viscously damped eigenmode that features latitudinal jets. We derive turbulent transport rates of momentum and heat, and provide them in analytic forms.  Such formulae, free of tunable model parameters, are tested against direct numerical simulations; the comparison shows good agreement. They improve upon prior quasi-linear or ``parasitic saturation” models containing a free parameter. Given model correspondences, we also extend this theory to heat and compositional transport for axisymmetric thermohaline instability-driven turbulence in certain regimes.
\end{abstract}


\section{Introduction} 
\label{intro}

\begin{figure*}
    \centering
    \includegraphics[width=0.96\textwidth]{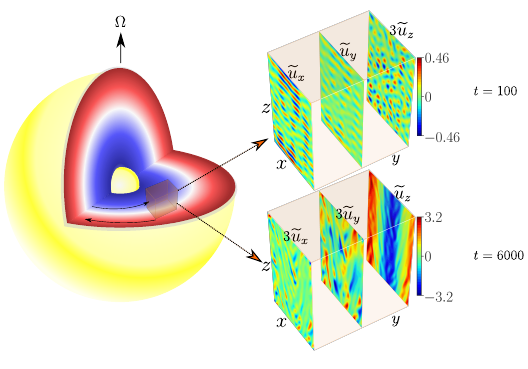}
    \caption{(Left) A schematic diagram of a differentially rotating star with a radial shear, gravity, and stable stratification. Such a system subject to the GSF instability is studied using a local Cartesian model. (Right) Snapshots of velocity components $\widetilde{u}_x(x,z)$, $\widetilde{u}_y(x,z)$, and $\widetilde{u}_z(x,z)$ from the axisymmetric GSF instability-driven turbulence; $x,y,$ and $z$ represent the local radial, azimuthal, and latitudinal directions. Though finger-like horizontal structures (as shown by, e.g., $\widetilde{u}_x$) grow the fastest in the linear phase ($t\mathrm{=}100$), strong latitudinal jets $\widetilde{u}_z$ are generated nonlinearly ($t\mathrm{=}6000$). The color bar for $t\mathrm{=}100$ is shared by $\widetilde{u}_x$, $\widetilde{u}_y$, and $3\widetilde{u}_z$; the color bar for $t\mathrm{=}6000$ is shared by $3\widetilde{u}_x$, $3\widetilde{u}_y$, and $\widetilde{u}_z$. The turbulent transport of angular momentum, e.g., $\langle \widetilde{u}_x \widetilde{u}_y\rangle$, is predicted in this paper using a jet-coupled turbulence closure.}
    \label{fig:fschematic}
\end{figure*}

Instability-driven turbulence is thought to play a major role in the  transport of angular momentum (AM), heat and composition in stellar and planetary interiors \citep[see, e.g.,][]{G2018,aerts2019,spruit2002,fuller2019}, as well as in astrophysical disks \citep[e.g.,][]{BH1998,Lesuretal2023}. Unfortunately, rates of turbulent transport are very challenging to predict theoretically, and the lack of reliable theories has hampered our understanding of the evolution of stellar and planetary internal rotations and structures. For example, the AM redistribution in red giant stars is currently poorly understood, and their core-envelope differential rotations inferred from asteroseismology have not been adequately explained \citep[e.g.,][]{B2012,RG2012,aerts2019}. The nearly solid-body rotation observed in the solar radiative interior also lacks a robust explanation \citep[e.g.,][]{GG2008,WM2011}.

Differential rotation is known to drive a variety of hydrodynamic (and hydromagnetic) instabilities. In this paper, we focus on modeling hydrodynamic instabilities of differential rotation in stellar and planetary radiative zones, and in particular on the Goldreich--Schubert--Fricke (GSF) instability\footnote{It has also been referred to as the ``Vertical Shear Instability” (VSI) in accretion disks \citep[e.g.,][]{UB1998,Nelson2013,BL2015,LP2018} and as ``inertial instability” enabled by thermal diffusion in stellar interiors \citep[][]{Park2020,Park2021}.}\citep{goldreich1967,fricke1968}. This is a double-diffusive centrifugal instability in which rapid thermal diffusion (relative to viscous momentum diffusion) enables instability by tempering the otherwise stabilising effects of buoyancy forces. Prior work has studied the linear and nonlinear properties of the instability, and the turbulence it drives \citep{KS1982,knobloch1982,K1991,Rashid2008,barker2019,barker2020,dymott2023}, but a reliable theory for the resulting turbulent transport is lacking. This means that the effects of the GSF instability on stellar rotational and chemical evolution have not been modeled in a self-consistent manner. Instead, one typically invokes unexplained ``additional viscosities” or models that contain free parameters. Such tunable parameters are intended to describe the effects of turbulence on AM transport for which adequate knowledge is lacking.

A fully analytic model containing no free parameters is derived here for the GSF instability-driven turbulence in $2.5$ dimensions ($2.5$-D), i.e., with all three components of velocity but varying spatially only in two dimensions. The predictions of our analytical model are in broad agreement with detailed numerical simulations of turbulence, driven by the axisymmetric ($2.5$-D) GSF instability at the equator of a star with radial differential rotation. Such model of the instability is, for certain diffusivity ratios and in $2.5$-D, formally and nonlinearly equivalent to the thermohaline, or salt-finger, instability that transports heat and chemical elements \citep{knobloch1982,barker2019}; thus, the turbulent transport arising from axisymmetric fingering convection is also described by our theory.

The structure of this paper is as follows. In \S~\ref{sec:model}, we present our model and methods of analysis. Nonlinear mode coupling and saturation diagnostics of the instability appear in \S~\ref{sec:theory}. Informed by such diagnostics, we present analytical formulae, without any free parameters, to model the turbulence and its transport properties in \S~\ref{sec:closure}. We discuss the astrophysical implications and conclude in \S~\ref{sec:discuss}. Details of the closure model are provided in the Appendices.

\section{The GSF instability and inertial-gravity waves}
\label{sec:model}

To study a basic mechanism of angular momentum transport in a differentially rotating star, we consider a local region inside the star near its equator, where the rotation can be split into a uniform or mean part---$\bm{\Omega}\mathrm{=}\Omega \hat{\bm{\mathrm{e}}}_z$, aligning with the local latitudinal axis $z$---and a non-uniform part due to the radial differential rotation. The latter is represented by a background linear shear flow $\bm{U}_0(x)\mathrm{=}-\mathcal{S}x\hat{\bm{\mathrm{e}}}_y$ where $x$ is the radial coordinate, $y$ is the azimuthal coordinate, and $\mathcal{S}=-d\Omega_\mathrm{Shell}(x)/d\ln x$ is the local radial-shear rate, with $\Omega_\mathrm{Shell}(x)$ representing the ``Shellular” rotation of the simplified star.  A uniform gravity field with $\bm{g} = -g\hat{\bm{\mathrm{e}}}_x$ is directed radially inward (Fig.~\ref{fig:fschematic}). A background radial temperature gradient $\nabla T_0$ then stratifies the fluid density radially, with a thermal expansion coefficient $\alpha$. In such a background state, any perturbations in velocity $\bm{u}$ and scaled temperature $\theta \mathrm{=} \alpha g T$ evolve \citep{barker2019} as 
\begin{subequations}
\begin{align} \label{eq:dtu}
    &D\bm{u}+\bm{u}\cdot\nabla\bm{U}_0 + 2 \bm{\Omega} \times \bm{u} = -\nabla p + \theta \hat{\bm{\mathrm{e}}}_x + \nu \nabla^2 \bm{u},\\ \label{eq:dtth}
    &D\theta+\bm{u}\cdot\nabla\Theta_0  = \kappa \nabla^2 \theta,\\
    &\nabla \cdot \bm{u} = 0,\\
    &D \equiv \partial_t + (\bm{u}+ \bm{U}_0) \cdot \nabla,
\end{align}
\end{subequations}
where the variables $p, \nu,$ and $\kappa$ are the fluid pressure (per unit density), the kinematic viscosity, and the thermal diffusivity, respectively. We also define the Prandtl number Pr$\mathrm{=}\nu/\kappa$. Because the GSF instability operates at length scales much smaller than a pressure scale height in stars, the local approximation is valid; in such a case, when the turbulence drives subsonic flows, the Boussinesq approximation is also appropriate \citep[][]{SV1960}. Assuming a uniform temperature gradient, a fluid element perturbed radially oscillates with a constant Brunt-V\"ais\"al\"a (buoyancy) frequency $\mathcal{N}$, where $\mathcal{N}^2 \hat{\bm{\mathrm{e}}}_x \mathrm{=} \nabla\Theta_0 \mathrm{=} \alpha g \nabla T_0$. 

Henceforth, we non-dimensionalize all variables using the characteristic rotation time scale $\Omega^{-1}$ and length scale $d$, with $d\mathrm{=}(\nu\kappa/\mathcal{N}^2)^{1/4}$, which is typically similar to the wavelengths of fastest-growing modes. Thus, $N \mathrm{=} \mathcal{N}/\Omega$ is the dimensionless buoyancy frequency and $S\mathrm{=}\mathcal{S}/\Omega$ the dimensionless shear rate (Rossby number). The GSF instability occurs in low-Pr fluids whenever $r\in [0,1]$, where $r=\mathrm{Pr} (1+ N^2 \kappa_\mathrm{ep}^{-2})/(\mathrm{Pr}-1)$, with $\kappa_\mathrm{ep}\mathrm{=}\sqrt{2(2-S)}$ representing the dimensionless epicyclic frequency \citep{barker2019}.

\subsection{Eigenmode analysis} \label{sec:eigmode}
A linear analysis of Eqs.~\eqref{eq:dtu}--\eqref{eq:dtth} for axisymmetric (uniform-in-$y$) perturbations yields a simple matrix equation, which upon Fourier-transforming becomes $\bm{L} \hat{\bm{X}}\mathrm{=}\gamma \hat{\bm{X}}$, where $\hat{\bm{X}}\mathrm{=}[\hat{u}_x, \hat{u}_y, \hat{u}_z, \hat{\theta}]^\mathrm{T}$, with ${}^\mathrm{T}$ as the transpose operation, is the state vector of spatially Fourier-transformed components at wavevector $\mathbf{k}\mathrm{=}(k_x, k_z)$; the matrix $\bm{L}$ is a linear operator, whose eigenvalues are the complex-valued growth rates $\gamma$. The size of $\bm{L}$ demands four linearly independent eigenvectors.  Because of the additional constraint $\nabla \mathrm{\cdot}\bm{u} \mathrm{=} 0$, the system has only three degrees of freedom at any given wavenumber---two components of velocity, and $\hat{\theta}$.  Hence, one eigenvector among the four eigenvectors does not satisfy $\nabla \mathrm{\cdot} \bm{u} \mathrm{=} 0$ and is rejected.  We confirm that this eigenvector is not excited within our incompressible Boussinesq simulations. One among the remaining three eigenvectors at a given wavevector becomes GSF-unstable [$\Real (\gamma) > 0$, where $\Real$ denotes the real part], whenever $r\in [0, 1)$. The remaining two eigenvectors are always stable, and their eigenvalues are complex conjugates of each other whenever they satisfy $\Imag(\gamma)\neq 0$, where $\Imag$ denotes the imaginary part; $\Imag(\gamma)$ corresponds to the frequency of inertial-gravity, or gravito-inertial, waves (IGWs), modified by the shear flow and damped by viscous and thermal diffusion.

The GSF instability grows dominantly via axisymmetric ($\partial_y \equiv0$) perturbations, therefore we focus upon the $(x,z)$-variations of the $3$-component velocity and temperature fields. The dispersion relation then is a simple cubic polynomial in $\gamma$ \citep{goldreich1967} as
\begin{equation}\label{eq:growthrateeqn}
    \gamma_\nu^2 \gamma_\kappa + \frac{\kappa_\mathrm{ep}^2 k_z^2}{k^2} \gamma_\kappa + \frac{N^2 k_z^2}{k^2} \gamma_\nu = 0,
\end{equation}
where $\gamma_\nu = \gamma + \nu k^2$ and $\gamma_\kappa = \gamma + \kappa k^2$. Equation~\eqref{eq:growthrateeqn} shows that, on the ($k_x, k_z$)-plane, the growth rate exhibits strong anisotropy: fluctuations with $k_x\mathrm{=}0$ (``elevator modes”) grow the fastest, whereas those with $k_z\mathrm{=}0$ are linearly stable. This observation is critical for the nonlinear saturation of the GSF instability because the anisotropy of the linear physics, in particular the $k_z\mathrm{=}0$ fluctuation, can impose its anisotropy on the nonlinear energy transfer, which is otherwise isotropic; such consequential effects have been found in various systems such as $3$-D Kelvin-Helmholtz instability \citep{tripathi2023b}, rotating  \citep{waleffe1993, smith1999} and stably stratified turbulence \citep{riley2000}, turbulence with an external magnetic field in astrophysical \citep{ng1996, du2023} and fusion plasmas \citep{biskamp1995, terry2004}.

Using the complete basis provided by the eigenvectors of the linear operator $\bm{L}$, we can decompose arbitrary incompressible fluctuation $\hat{\bm{X}}_{\mathrm{arb}}$ with $k_z\mathrm{\neq} 0$ as 
$\hat{\bm{X}}_{\mathrm{arb}} \mathrm{=} \sum_{j=1}^{3} \beta_j \hat{X}_j$, where $\beta_j$ is the amplitude of the $j^\mathrm{th}$ eigenvector $\hat{X}_j$; we reserve $j\mathrm{=}1$ for the GSF-unstable modes, and $j\mathrm{=}2,3$ for the IGWs that are always linearly stable in this study. More compactly, $\hat{\bm{X}}_{\mathrm{arb}} \mathrm{=} \bm{E} \bm{\beta}$, where $\bm{\beta}$ is a (column-)vector of mode amplitudes and $\bm{E}$ is an eigenvector matrix, whose $j^\mathrm{th}$ column is $\hat{X}_j$. Thus, $\bm{\beta}\mathrm{=} \bm{E}^{-1} \hat{\bm{X}}_{\mathrm{arb}} $. 

For the $k_z\mathrm{=}0$ modes, $\bm{E}$ turns out to be an identity matrix, meaning that the three components of velocity, and the temperature, individually form eigenvectors.  In what follows, we therefore decompose an arbitrary fluctuation with $k_z\mathrm{=}0$ into $\hat{\bm{X}}_{\mathrm{arb}}^\mathrm{T} = \X[1,0,0,0] + \Y[0,1,0,0]+\Z[0,0,1,0] +\T[0,0,0,1]$, where the amplitudes of the eigenvectors are denoted by $\X,\Y,\Z$, and $\T$. We reserve the $\bm{\beta}$-notation above for the amplitudes of eigenvectors with $k_z\mathrm{\neq}0$.

\subsection{Initial value problem}

We perform an ensemble of direct numerical simulations of Eqs.~\eqref{eq:dtu} and \eqref{eq:dtth}, by seeding a low-amplitude solenoidal random noise to $\bm{u}$, in a box of size $(L_x, L_z) \mathrm{=} (100,100)$. To obtain numerically converged results, a spatial resolution of up to $512^2$ grid points is used in the pseudo-spectral solver SNOOPY \citep{lesur2005, barker2019}.

\begin{figure*}
    \centering
    \includegraphics[width=0.65\textwidth]{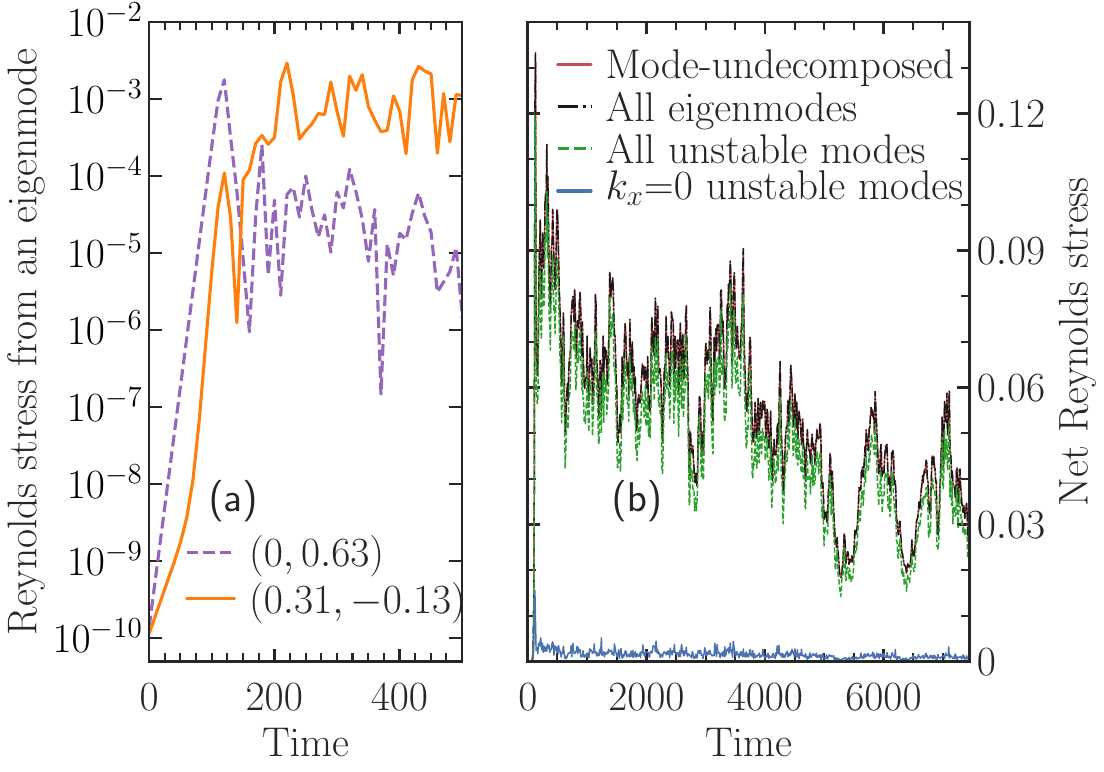}
    \caption{(a) Comparison of momentum transport ($\mathrm{Reynolds\ Stress}\mathrm{=}2|\beta_1|^2 \mathrm{Re}\left[ \hat{u}_{x,1} \hat{u}_{y,1}^\ast\right]$) driven by an unstable mode at the linearly fastest-growing wavenumber $\mathbf{k}=(0,0.63)$, and by an unstable mode at $\mathbf{k}=(0.31,-0.13)$, the wavenumber that has the largest contribution to the momentum transport in the nonlinear phase. (b) Eigenmode decomposition of net Reynolds stress $\langle \widetilde{u}_x \widetilde{u}_y \rangle$ in nonlinear simulation of the GSF instability-driven turbulence, showing that the transport due to mode-undecomposed fluctuations (red curve) and mode-decomposed all eigenmodes (black curve) agree to machine precision. Transport is almost entirely ($88\%$) due to the unstable modes (green curve); the sum of fastest-growing unstable modes at $k_x\mathrm{=}0$, however, contributes negligibly ($3\%$) to the transport (blue curve). Simulation parameters used are $S\mathrm{=}2.1, N^2\mathrm{=}10$ and Pr$\mathrm{=}0.01$.}
    \label{fig:f1}
\end{figure*}

To determine the contribution of each eigenmode in, for example, the turbulent momentum transport, we decompose the turbulent stress as: $\langle \widetilde{u}_x \widetilde{u}_y \rangle \mathrm{=} \sum_{k_x, k_z} \sum_{m,n} 2\mathrm{Re}\left[\beta_m  \hat{u}_{x,m} \beta_n^\ast\hat{u}_{y,n}^\ast\right]$, where $\langle\cdot\rangle$ is an $(x,z)$-averaging operation; $m$ and $n$ are summed from $1$ to $3$, corresponding to three excited eigenmodes at every wavenumber $\bm{k}$; the amplitude $\beta_m$ and the $x$-component of the velocity $\hat{u}_{x,m}$ correspond to the $m^\mathrm{th}$ eigenvector at $\bm{k}$; and likewise for $\beta_n$ and $\hat{u}_{y,n}$; the operation $^\ast$ denotes complex conjugation.  Using such a decomposition, we obtain the contribution of an unstable mode at $\bm{k}$ to the momentum transport rate, which is $2|\beta_1|^2 \mathrm{Re}\left[ \hat{u}_{x,1} \hat{u}_{y,1}^\ast\right]$. This decomposition is performed for every wavenumber, hence allowing us to trace evolution of transport contributions due to individual unstable modes [see Fig.~\ref{fig:f1}(a)].

The summed contributions of all eigenvectors from all wavenumbers reproduce, to machine precision, the total transport rates found in the simulation before performing mode decomposition, as we show in Fig.~\ref{fig:f1}(b).  The contributions of the unstable modes are also compared across different wavenumber sums.  Almost identical results are found for heat transport (not shown). The unstable modes from the linearly fastest-growing wavenumber branch $k_x\mathrm{=}0$ transport significantly less momentum than the other wavenumbers with $k_x\mathrm{\neq}0$. This is our first surprising result, and it challenges predictions of turbulent transport that rely on an unstable mode at the fastest-growing wavenumber alone \citep[e.g.,][]{radko2012, brown2013, barker2019}.  This finding also instructs us to investigate nonlinear couplings between eigenmodes to understand the instability-saturation mechanism.

\begin{figure*}
    \centering
    \includegraphics[width=0.62\textwidth]{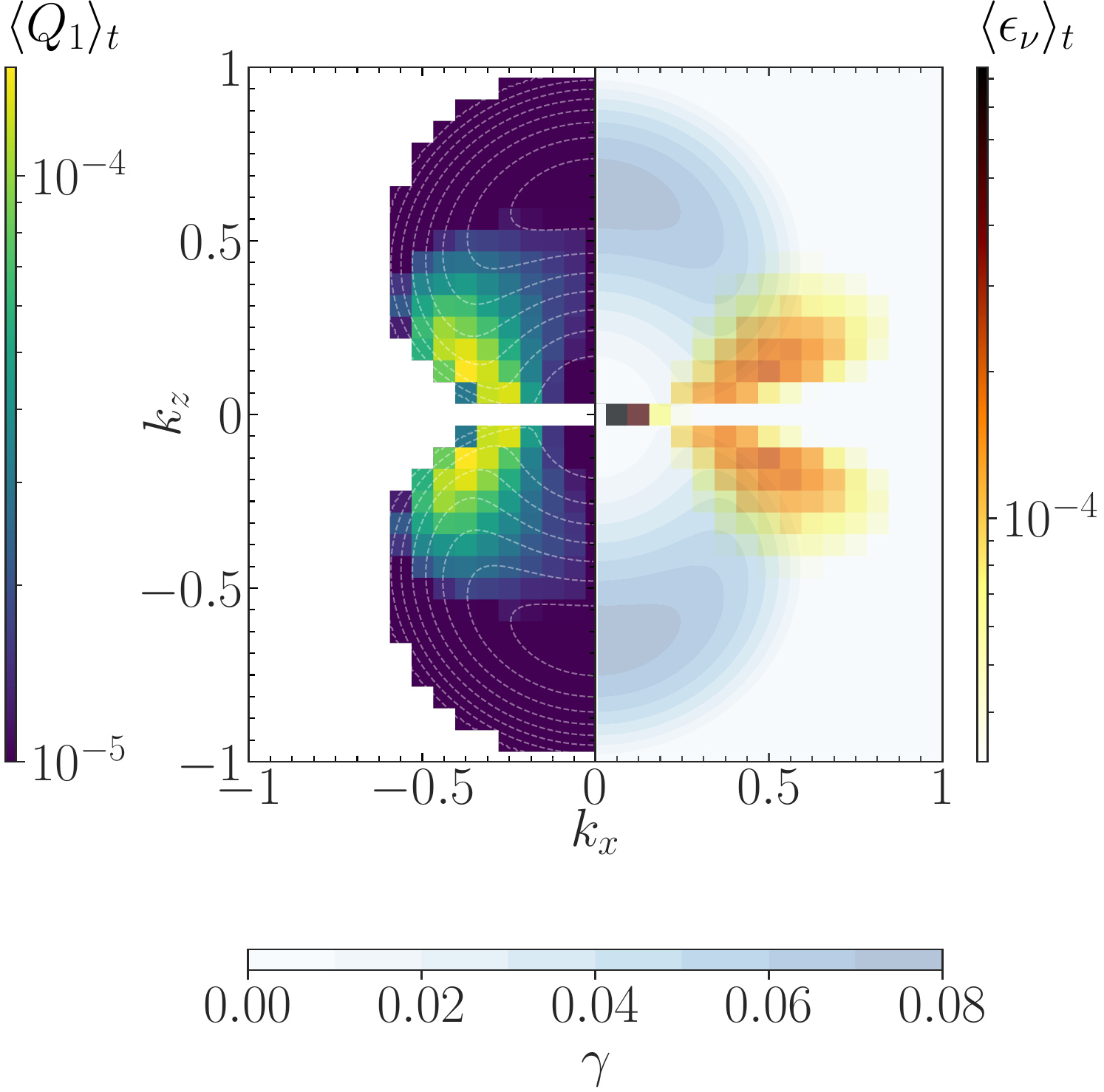}
    \caption{Spectra of linear and nonlinear-saturation  properties of the GSF instability. On the negative-$k_x$ domain, the colored square boxes (yellow-green-purple) display time-averaged energy extraction rates $\langle Q_1 \rangle_t$ by unstable modes from the mean gradients in a nonlinear simulation. On the positive-$k_x$ domain, colored square boxes (black-red-yellow) show the time-averaged viscous dissipation rates, which are pronounced at low $k_z$. Over the entire $(k_x, k_z)$-plane, the non-square filled and line contours show the growth rates $\gamma$ of unstable modes, with white dashed contour lines on the negative-$k_x$ domain and with bluish filled contours on the positive-$k_x$ domain. The fastest-growing mode resides at around $\mathbf{k}=(0,0.63)$. The simulation parameters are $S\mathrm{=}2.1, N^2\mathrm{=}10$ and Pr$\mathrm{=}0.01$.}
    \label{fig:f2}
\end{figure*}

\section{Nonlinear saturation by coupling to latitudinal flow}
\label{sec:theory}

\subsection{Mode-amplitude evolution}

To analyze nonlinear mode couplings, Eqs.~\eqref{eq:dtu}--\eqref{eq:dtth} are first spatially Fourier-transformed: $\partial_t \hat{\bm{X}} = \bm{L} \hat{\bm{X}} + \sum_{\bm{k'},\bm{k''}} \bm{N}(\hat{\bm{X}}', \hat{\bm{X}}'')$, where $\hat{\bm{X}},\hat{\bm{X}}'$, and $\hat{\bm{X}}''$ are state vectors at $\bm{k},\bm{k}'$, and $\bm{k}''$, respectively, satisfying $\bm{k}=\bm{k}'+\bm{k}''$. Then, following Sec.~\ref{sec:eigmode} we substitute $\hat{\bm{X}} \mathrm{=} \bm{E} \bm{\beta}$, and likewise for $\hat{\bm{X}}'$ and $\hat{\bm{X}}''$. We multiply the obtained equation with $\bm{E}^{-1}$ and take the $j^\mathrm{th}$ row of the resulting equation.  This process yields an evolution equation for the $j^\mathrm{th}$ eigenmode at $\bm{k}$. Such an evolution equation for mode amplitude $\beta_j$ for $k_z\mathrm{\neq}0$ is different from that of the mode amplitude $F \in \{\X,\Y,\Z,\T\}$ for $k_z\mathrm{=}0$, although both are coupled and nonlinear:
\begin{subequations}
\begin{align}\label{eq:betaevoln}
    \partial_t \beta_j &= \gamma_j \beta_j + \sum_{\substack{\bm{k}',m,n}} C_{jmn}^{(\bm{k},\bm{k}')}  \beta_m'\beta_n'' \nonumber \\
    &\hspace{0.6cm}+  \sum_{\substack{\bm{k}',F,n: k_z'\mathrm{=}0\\ F \in \{\X,\Y,\Z,\T\}}} \left[ C_{j F n}^{(\bm{k},\bm{k}')} + C_{j n F}^{(\bm{k},\bm{k}'')}  \right] F' \beta_n'',\\ \label{eq:Fevoln}
    \partial_t F &= -\gamma_F F + \sum_{\bm{k}',m,n: k_z\mathrm{=}0} C_{F m n}^{(\bm{k},\bm{k}')} \beta_m'\beta_n'',
\end{align}
\end{subequations}
where $\gamma_j$ is the complex-valued growth rate for the $j^\mathrm{th}$ eigenmode with $k_z\mathrm{\neq}0$; the real-valued damping rate $\gamma_F$ is $\gamma_\X$ when $F$ is replaced with $\X$ in Eq.~\eqref{eq:Fevoln}; likewise for the replacement of $F$ with $\Y, \Z,$ and $\T$; we note that $\gamma_\X = \gamma_\Y = \gamma_\Z = \nu k_x^2$, and $\gamma_\T = \kappa k_x^2$.  The nonlinear-coupling coefficient, for example, $C_{jmn}^{(\bm{k},\bm{k}')}$ measures the overlap of eigenmodes $m$ with $\bm{k}'$,  $n$ with $\bm{k}''$, and $j$ with $\bm{k}$.  Such a mode coupling coefficient is found by applying $\bm{E}^{-1}$ to the (column) vector of $\bm{N}(\hat{\bm{X}}_m', \hat{\bm{X}}_n'')$, a process that incorporates all the nonlinearities of the system, thus making $C_{jmn}^{(\bm{k},\bm{k}')}$ ideal for a comprehensive instability-saturation analysis.

\begin{figure*}
    \centering
    \includegraphics[width=1\textwidth]{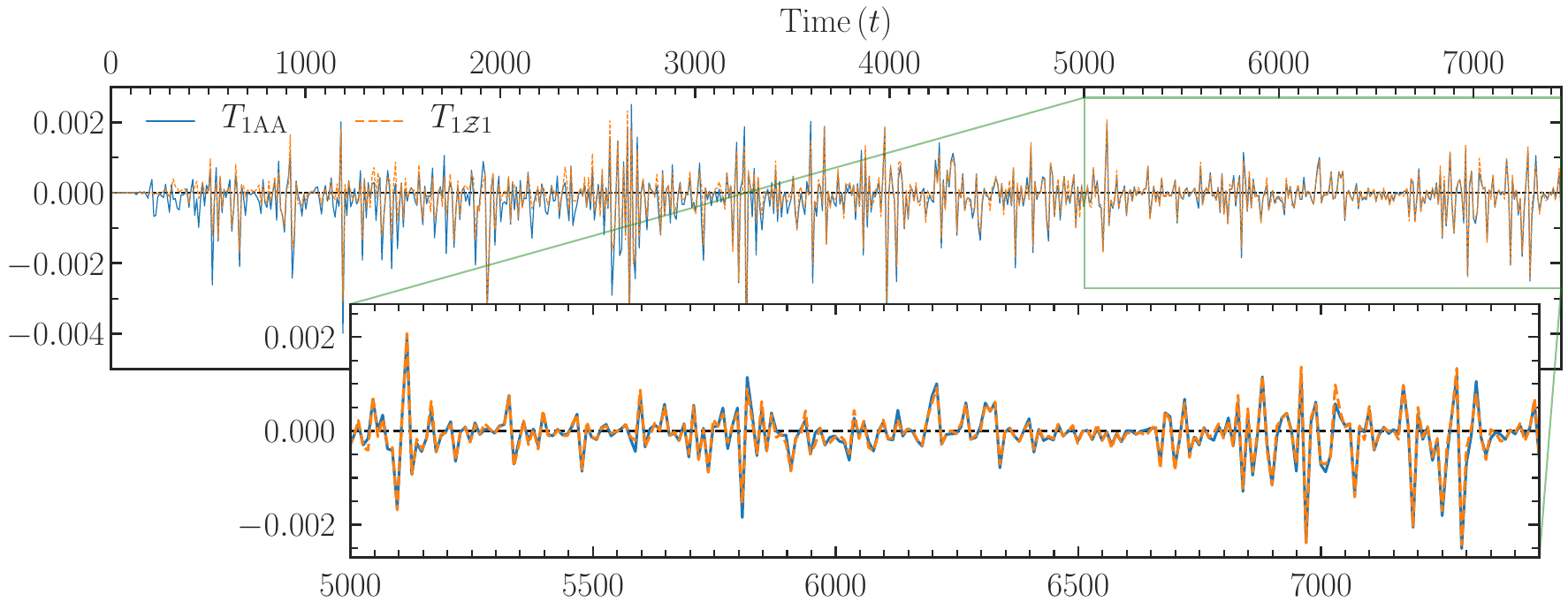}
    \caption{Time evolution of the total nonlinear energy transfer $T_{1\mathrm{AA}}$ to an unstable eigenmode at a wavenumber where the spectrum of $\langle \tilde{u}_x\tilde{u}_y\rangle$  peaks. $T_{1\Z 1}$ is the energy transfer to the same unstable mode via interactions between the $z$-component of velocity ($\Z$) with wavenumbers $k_z\mathrm{=}0$, and the other unstable modes. Comparison  of two transfer functions reveals that the dominant triad involves a latitudinal flow and two unstable modes. The simulation parameters are $S\mathrm{=}2.1, N^2\mathrm{=}10$ and Pr$\mathrm{=}0.01$.}
    \label{fig:f3}
\end{figure*}

On the right-hand side of Eq.~\eqref{eq:betaevoln}, the second term is the nonlinear coupling between eigenmodes with $k_z'\mathrm{\neq}0$ and $k_z''\mathrm{\neq}0$ (hence the two $\beta$s), and the third term, with an $F$ and a $\beta$, is the nonlinear coupling between eigenmodes with $k_z'\mathrm{=}0$ and $k_z''\mathrm{\neq}0$ (see the last paragraph of Sec.~\ref{sec:eigmode}). Equations~\eqref{eq:betaevoln} and \eqref{eq:Fevoln} have the same number of degrees of freedom as the original nonlinear equations in physical space, Eqs.~\eqref{eq:dtu}--\eqref{eq:dtth}. These systems are completely equivalent, but one represents dynamics in physical space and another in eigenmode space.

\subsection{Mode-energy evolution}

The energy evolution equation for each eigenmode can now be derived by multiplying Eq.~\eqref{eq:betaevoln} by $\beta_j^\ast$ 
and adding the complex conjugate of the resulting equation to arrive at
\begin{equation}\label{eq:dtbetasqrdmain}
    \partial_t |\beta_j|^2 = Q_j + T_{j\mathrm{A}\mathrm{A}},
\end{equation}
where $Q_j\mathrm{=}2\Real \gamma_j |\beta_j|^2$ is the linear energy transfer rate to $\bm{k}$ from the mean gradients, and $T_{j\mathrm{A}\mathrm{A}}$ is the total nonlinear energy transfer to the $j^\mathrm{th}$ eigenmode from \textit{all} possible nonlinear interactions; Eq.~\eqref{eq:energyevoln1}. We now show the spectrum of time-averaged $Q_1$, along with that of the growth rate and time-averaged viscous dissipation rate $\epsilon_\nu$ in Fig.~\ref{fig:f2}. 

From $T_{j\mathrm{A}\mathrm{A}}$ in Eq.~\eqref{eq:dtbetasqrdmain}, we separate out the nonlinear transfer $T_{1\mathrm{\Z}1}$ in a triad that involves a latitudinal flow $\Z$ at $k_z'\mathrm{=}0$ and two GSF-unstable modes ($j\mathrm{=}1$) at $k_z'\mathrm{\neq} 0$:
\begin{equation}
    T_{1\mathrm{\Z}1}= \sum_{\bm{k}':k_z'=0} 2\Real \Bigg\{  \left[ C_{1 \Z 1}^{(\bm{k},\bm{k}')} + C_{11 \Z}^{(\bm{k},\bm{k}'')}  \right] \Z' \beta_1'' \beta_1^\ast \Bigg\} \label{eq:energyevolnY}.
\end{equation}

We now compare $T_{1 \Z 1}$ with $T_{1\mathrm{A}\mathrm{A}}$ in Fig.~\ref{fig:f3}, for a GSF-unstable mode with a wavenumber that contributes the largest to the momentum transport. Repeating this transfer analysis at different wavenumbers produces similar results. The two transfers are nearly identical, which confirms the conjecture \citep{barker2019} that, in the fully nonlinear phase, the GSF instability saturates via the formation of strong latitudinal jets or flows. Such flows are $z$-directed, although with no $z$-variation, and primarily have a wavenumber $k_x\mathrm{=}2\pi/L_x$; see Fig.~\ref{fig:fschematic}, right-column snapshot at $t\mathrm{=}6000$. These flows generalize to meridional circulation in stars, and resemble zonal jets in planetary atmospheres and fusion systems \citep{terry2019}. Flows with $k_z\mathrm{=}0$ are, however, linearly stable to the GSF instability, and, thus, must necessarily be excited nonlinearly by the interactions between the GSF-unstable modes. This energy received is then viscously damped at $k_z\mathrm{=}0$ and at low $k_z$, as seen in Fig.~\ref{fig:f2}. To sum up, the mean shear flow, destabilized by the thermal diffusion, lends energy to the fluctuations via the GSF-unstable modes, which saturate by exciting $k_z\mathrm{=}0$ latitudinal flows to a significant level. Such flows then viscously dissipate the turbulent energy. This is the saturation mechanism of the axisymmetric GSF instability found here.

\section{Angular momentum transport model}\label{sec:closure}

The findings shown so far are sufficient to build a statistical closure model, with no free parameters, and thus with predictive power.

Equation~\eqref{eq:betaevoln} has a quadratic nonlinearity, hence evolutionary equations for mode energy contain triplet interactions [e.g., Eq.~\eqref{eq:energyevolnY}]. To determine the evolving triplet interaction terms, one can derive an equation with quadruplet interactions [Eq.~\eqref{eq:b11}], and so on.  To truncate this never-ending hierarchy (the so-called ``turbulence closure problem”), we invoke a standard turbulence closure, the Eddy-Damped Quasi-Normal Markovian (EDQNM) approximation (see, e.g., \citealt{orszag1970, terry2018, hegna2018, terry2021, pueschel2021, li2021, li2023}), that truncates the hierarchy at fourth-order cumulants of the fluctuations, thereby assuming that the statistics for the mode amplitudes are close to Gaussian. The resulting equation, however, is still nonlinear and daunting. But when a latitudinal flow $\Z$, with $k_z\mathrm{=}0$, dominates the nonlinear coupling, the complexity of the equation is significantly reduced \citep{terry2018}.

\subsection{An outline of the Closure Model}
We illustrate here the key steps involved to explain most simply our closure model (by omitting details and treating all variables as real). First, we observe that Eq.~\eqref{eq:betaevoln} has the structure: 
\begin{equation}
\partial_t \beta = ...\beta + ...\beta \beta + ... \beta \X + ... \beta \Y + ... \beta \Z + ... \beta \T\, ,
\end{equation} where the mode amplitudes are explicitly shown, and the dots ($...$) represent terms such as the linear growth rate and the nonlinear coupling coefficients. One can then obtain evolution equation for second-order correlator as $\partial_t (\beta\beta) = ...\beta\beta +... \beta \Z \beta$, where the other nonlinear terms, e.g., $\beta\beta\beta$ and $\beta\beta\X$, have been dropped because the nonlinear energy transfer is almost entirely dominated by $\beta \Z \beta$---the latitudinal-flow coupling (Fig.~\ref{fig:f3}). Since $\beta \Z \beta$ also evolves, one can similarly obtain evolution equation for third-order correlator as $\partial_t (\beta\Z\beta) = ...\beta\Z\beta +... \beta\beta \Z\Z $. The closure solution then yields a relation $\beta\Z\beta =... \beta\beta \Z\Z $.  Although useful later, this relation does not predict the mode amplitude $\beta$, needed for the turbulent transport prediction.

To predict the mode amplitude $\beta$, we consider the latitudinal-flow evolution equation, $\partial_t\Z = ...\Z + ...\beta\beta$, and derive $\partial_t(\Z\Z) = ...\Z\Z + ...\beta\Z\beta$. Then, $\beta\Z\beta$ can be replaced with a product of four amplitudes using the closure solution in the previous paragraph. One thus obtains $\partial_t(\Z\Z) = ...\Z\Z + ...\beta\beta\Z\Z$. In quasi-stationary turbulence, $\partial_t \sim 0$, and thus $\beta\beta\Z\Z =...\Z\Z$. The EDQNM closure allows writing a fourth-order correlator $\beta\beta\Z\Z$ as a sum of products of second-order correlators, such as $|\beta|^2 |\Z|^2$. Then, cancelling $|\Z|^2$ from both sides of $\beta\beta\Z\Z =...\Z\Z$, one predicts the saturated mode amplitude (or energy): $\beta\beta =...\,$. Using this, one can make predictions for turbulent transport rates as we shall show in the next subsection.

\subsection{Detailed Closure Model}

To make quantitative predictions for transport, we take the EDQNM-closed evolution equation for the latitudinal flow energy [see Appendix~B, Eq.~\eqref{eq:dtz2appendix}],
\begin{equation}\label{eq:dtz2}
\begin{aligned}
\partial_t |\Z|^2/2 
=  &-\gamma_\Z |\Z|^2 \\
&+ |\Z|^2  \sum_{\bm{k}'}|\beta_1'|^2 \,   \Real \Bigg. \left[-\tau_{11\Z} \dvec C_{\Z 11}^{(\bm{k},\bm{k}')} \dvec C_{1\Z 1}^{(\bm{k}'',\bm{k})}  \right],
\end{aligned}
\end{equation}
where, on the right-hand side, the second term contains a product of four amplitudes, but notably with $|\Z|^2$ that also appears in the first term of the right-hand side. In Eq.~\eqref{eq:dtz2}, 
\begin{equation}
\tau_{11\Z} = (\gamma_1' +\gamma_1'' -\gamma_\Z^\ast)^{-1},
\end{equation}
is the three-wave interaction time found from the EDQNM closure, and $\dvec C_{lmn}^{(\bm{p},\bm{q})} = C_{lmn}^{(\bm{p},\bm{q})} + C_{lnm}^{(\bm{p},\bm{p}-\bm{q})}$ is the symmetrized coupling coefficient. In quasi-stationary turbulence, $\partial_t \sim 0$, and thus the linear and nonlinear terms must balance. First, for simplicity, we consider a latitudinal flow at $(k_x, 0)$ that is driven by two unstable modes at $(k_x',k_z')$ and $(k_x\mathrm{-}k_x', \mathrm{-}k_z')$; then, using \eqref{eq:dtz2},  $|\beta_1'|^2 = \gamma_\Z \left(\Real [-\tau_{11\Z} \dvec C_{\Z 11}^{(\bm{k},\bm{k}')} \dvec C_{1\Z 1}^{(\bm{k}'',\bm{k})} ]\right)^{-1}$. A more general expression for $|\beta_1'|^2 $ is found by using a standard Markovian assumption \citep{terry2021}: $|\beta_1'|^2$ is more weakly dependent on wavenumbers than the other factors in Eq.~\eqref{eq:dtz2} arising from the coupling coefficients and $\tau_{11\Z}$. Such a consideration provides an expression for nonlinearly saturated squared-mode-amplitude
\begin{equation}
    |\beta'|^2 = \gamma_\Z \times {\left(\frac{\gamma}{k^2}\right)}_{\mathrm{Closure}},
\end{equation}
with
\begin{equation}\label{eq:gk2}
    {\left(\frac{\gamma}{k^2}\right)}_{\mathrm{Closure}} = \frac{1}{ |\sum_{\bm{k'}}\tau_{11\Z}\,  \Real [-\dvec C_{\Z 11}^{(\bm{k},\bm{k}')} \dvec C_{1\Z 1}^{(\bm{k}'',\bm{k})} ]| },
\end{equation}
where we note that the coupling coefficients scale linearly with wavenumbers, and $\tau_{11\Z}$ is the inverse of the sum of three growth rates of eigenmodes in a triad. 

The growth rates in $\tau_{11\Z}$ should, in principle, also have amplitude-dependent eddy-damping rates as they become non-negligible, for example, in homogeneous isotropic fluid turbulence; however, when waves or instabilities exist, and when the turbulent transport spectrum is dominated by low wavenumbers, as in this study, $\tau_{11\Z}$ is approximated by using the linear growth rates \citep{terry2018, terry2021}. Using such, one identifies that the triplet interaction time $\tau_{11\Z}$ is maximal when the triad involves a latitudinal flow ($\Z$) and two GSF-unstable modes ($j\mathrm{=}1$).  Shorter triplet interaction times $\tau_{12\Z}$ are expected for triads with, for example, the latitudinal flow, an unstable mode, and a strongly damped IGW ($j\mathrm{=}2$), as such an interaction lowers $\tau_{12\Z}$ via both the frequency and damping rate of the IGW. The largest interaction time $\tau_{11Z}$ dominates saturation.

The radial turbulent transport of angular momentum is measured by $ \langle \tilde{u}_x \tilde{u}_y \rangle  \mathrm{\approx} \sum_{\bm{k}'''}  \hat{u}_{x,1}''' \hat{u}_{y,1}'''^\ast |\beta_1'''|^2$,
where $\beta_1'''$ is the unstable-mode amplitude at $\bm{k}'''$ over which the summation is applied. Then, using $|\beta_1'''|^2 $ from the above paragraph,
\begin{equation} \label{eq:uxuy}
    \langle \tilde{u}_x \tilde{u}_y \rangle_\mathrm{Closure}  = {\left(\frac{\gamma}{k^2}\right)}_{\mathrm{Closure}}\times \gamma_\Z
 \sum_{\bm{k}'''}   \hat{u}_{x,1}''' \hat{u}_{y,1}'''^\ast.
\end{equation}
The $y$-component $\hat{u}_{y,1}'''$ of velocity of the unstable eigenvector can be replaced with, e.g., its temperature perturbation $\hat{\theta}_{1}'''$ to predict the turbulent heat flux $\langle \tilde{u}_x \tilde{\theta} \rangle$. 

In our simulations with radial differential rotation, the latitudinal momentum flux is much lower than the radial flux, and,  when time-averaged, it is nearly null.

\subsection{Tests of theoretical predictions}

\begin{figure*}
    \centering
    \includegraphics[width=0.9\textwidth]{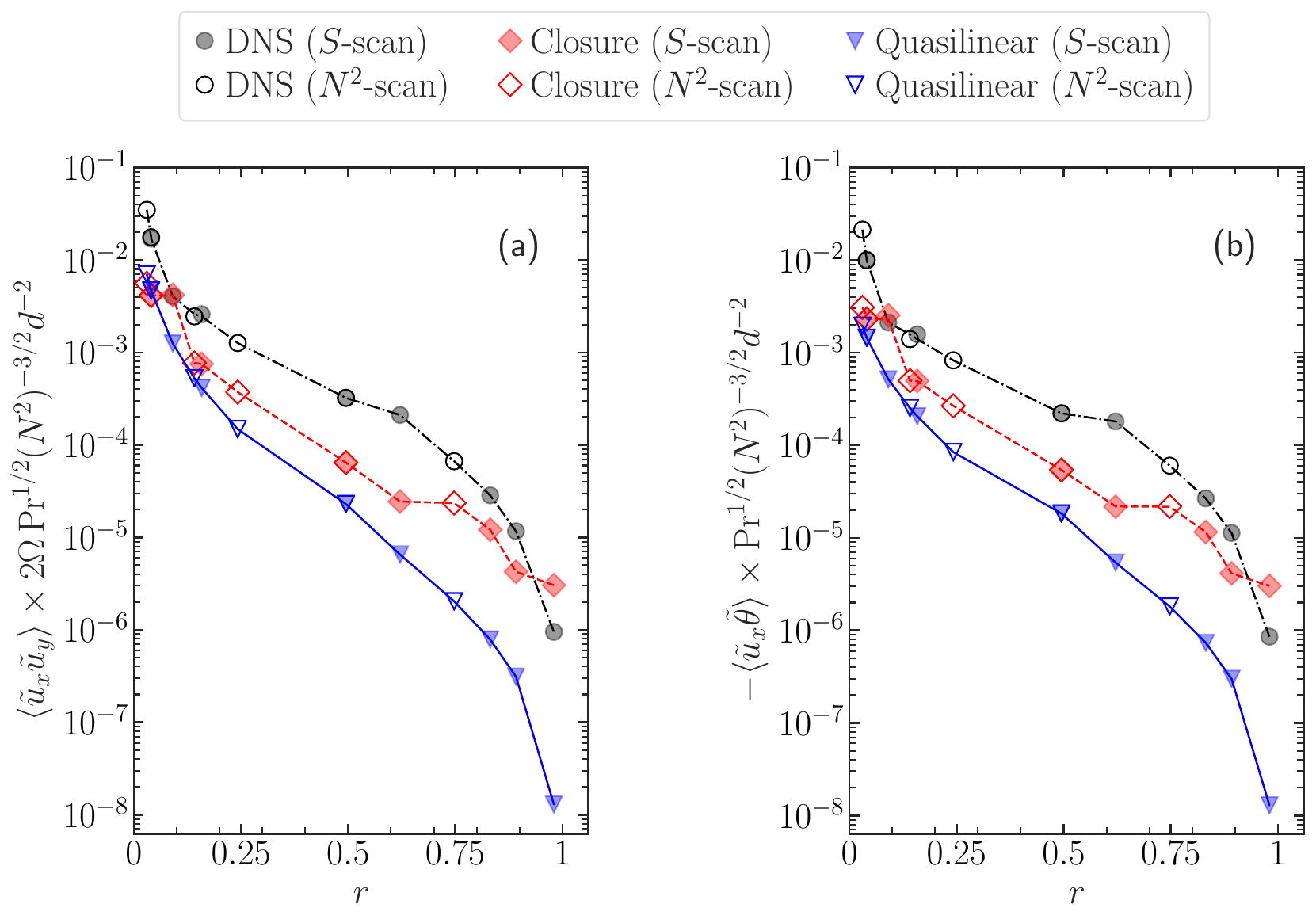}
    \caption{Tests of predictions of our closure model (red diamond) and a quasilinear-type, parasitic-saturation model (blue inverted triangle) against direct numerical simulations (DNS, black/gray circle). Variations of momentum transport rates are shown in (a); the filled markers correspond to the cases where the shear parameter $S$ is varied ($N^2\mathrm{=}10$); the unfilled markers correspond to the cases where the squared Brunt-V\"ais\"al\"a frequency $N^2$ is varied  ($S\mathrm{=}2.1$). Both $S$- and $N^2$-scan results collapse onto a single master curve, when $\langle \tilde{u}_x \tilde{u}_y\rangle$ is scaled by a factor shown on the $y$-axis that transforms the governing equations of the GSF instability studied here to depend on only two dimensionless parameters $(r,\mathrm{Pr})$.  The GSF instability operates when $r\in [0,1)$ and $\mathrm{Pr}\mathrm{<}1$ ($\mathrm{Pr}\mathrm{=}0.01$ is chosen). The shown $y$-axis is precisely an expression for the chemical transport rate for the thermohaline instability [see Eq.~\eqref{eq:figcaptionm}]. Heat transport rates, shown in (b), display nearly identical trends; [see Eq.~\eqref{eq:figcaptionT} for the scaling factor]. The closure prediction agrees with full DNS better than the quasilinear prediction over the scanned range of parameters.}
    \label{fig:f4}
\end{figure*}

A simple quasilinear model of the GSF-instability saturation was recently proposed \citep{barker2019, barker2020} by assuming that a secondary ``parasitic” instability feeds on the primary GSF-unstable mode. Such an assumption, also called ``parasitic saturation mechanism” \citep{goodman1994,radko2012,brown2013,harrington2019,fraser2023}, is based on a single primary mode at the fastest-growing wavenumber $\mathbf{k}'''$, which predicts the transport rate  
\begin{equation}\label{eq:uxuyQL}
\langle \widetilde{u}_x \widetilde{u}_y \rangle_{\mathrm{QL}} = \frac{ \gamma_1'''^2  }{k_z'''^2} \hat{u}_{x,1}''' \hat{u}_{y,1}'''^\ast f^2(\mathbf{k}'''),
\end{equation}
whose form is made manifestly similar to Eq.~\eqref{eq:uxuy}; the factor $f(\mathbf{k}''')$, which is evaluated at $\mathbf{k}'''\mathrm{=}(0, k_z''')$, is the normalization factor of eigenmodes. Here, $\gamma_\nu'''=\gamma_1'''+\nu k'''^2$. To find $\langle \widetilde{u}_x \widetilde{\theta} \rangle_{\mathrm{QL}}$, one can replace $\hat{u}_{y,1}'''$ on the right-hand side of Eq.~\eqref{eq:uxuyQL} with $\theta_1'''$. 

Predictions of Eqs.~\eqref{eq:uxuy}~and~\eqref{eq:uxuyQL} are compared against transport rates from direct numerical simulations in Figs.~\ref{fig:f4}. Significant improvement in both the momentum and heat transport predictions is observed with the statistical closure model. The orders-of-magnitude variation in transport rates is captured by the closure model.

In Fig.~\ref{fig:f4}, for smaller values of $r$, though the transport rates of the closure and the quasilinear models are similar, we emphasize that this similarity is merely accidental: the physics the two models incorporate is very different. The assumptions of the closure model are supported by detailed numerical evidence (Figs.~\ref{fig:f1} and \ref{fig:f2}), including that of the dominant three-wave coupling between two unstable modes and a latitudinal jet (Fig.~\ref{fig:f3}). No jet physics is considered in the quasilinear model. The quasilinear model predicts transport rates based on only one fastest growing wavenumber $k_z$, with $k_x\mathrm{=}0$, which Fig.~\ref{fig:f1} shows is inadequate. Hence, predictions of the quasilinear model that tend to reproduce the data-validated closure model predictions are at best a fortuitous coincidence, occurring in a very limited parameter regime.

\subsection{Impact of the new model in astrophysics}

Since stellar interiors typically have extreme parameters such as $\mathrm{Pr}\lesssim 10^{-6}$, current and anticipated near-future computational resources are insufficient to permit direct numerical simulations of realistic turbulence in them. In the face of such a challenge, progress can be made by developing analytical theories, informed and tested by numerical simulations at more accessible parameters. Thus, we now employ our analytical theory to extrapolate and make predictions for realistic astrophysical parameters. To achieve this, we derive fully analytic expressions for all elements of the closure model, assuming that the coupling of two GSF-unstable modes with the latitudinal jet remains dominant; see Appendices A and B. 

\begin{figure*}
    \centering
    \includegraphics[width=0.85\textwidth]{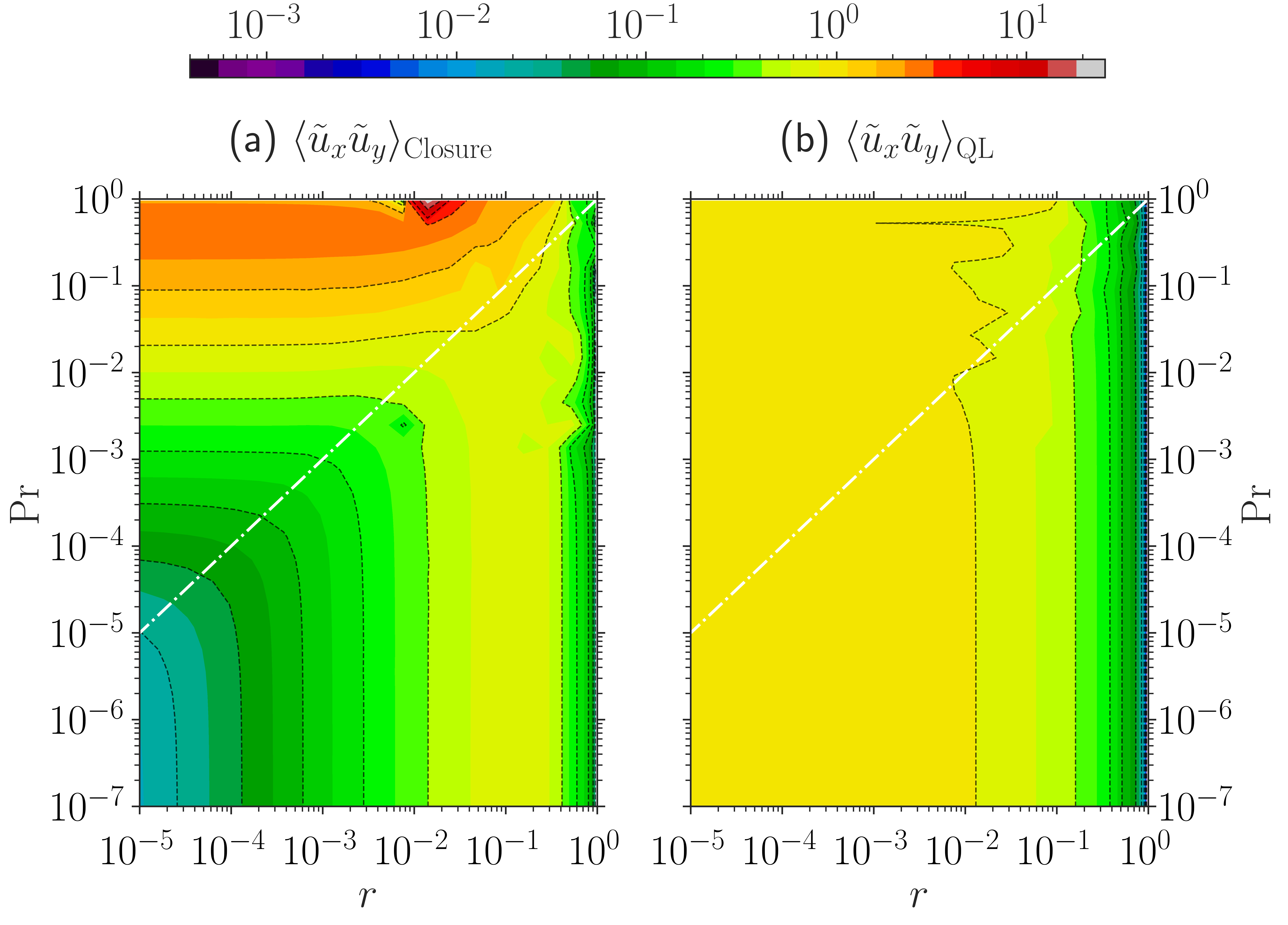}
    \caption{Predictions of (a) closure model and (b) quasilinear (QL) model for turbulent momentum transport.  The QL model, largely independent of Pr, fails to reproduce the behavior of the closure model. For Pr closer to $1$, the closure model predicts that the transport rate increases with decreasing $r$---until an asymptotically large transport is attained. For $\mathrm{Pr}\ll r$, the transport first increases and then decreases with $r$, in contrast to the QL prediction. The white dashed-dotted line, with a unit slope, separates distinct regimes of $\mathrm{Pr}< r$ and $\mathrm{Pr}> r$ found in (a).
    }
    \label{fig:f5}
\end{figure*}

We then compare predictions of the closure model with those of the quasilinear (QL) model, over a wide range of parameters $\mathrm{Pr}\approx 10^{-7}\textrm{--}1$ and $r\approx 10^{-5}\textrm{--}1$ in Fig.~\ref{fig:f5}. Noting that $N^2 \mathrm{=} 2(S-2)\left[1+r(\mathrm{Pr}^{-1}-1)\right]$, these scans span $N^2\lesssim 19 \times 10^6$ (in terms of $\Omega^2$); this ratio is typically around 1 million for the Sun \citep{christ1996}. In Fig.~\ref{fig:f5}, with $S\mathrm{=}3$ (in terms of $\Omega$), the Richardson number is as large as $\approx 2\times 10^{6}$. More extreme parameters can be easily and quickly scanned with the analytic formula we have derived.

Now we predict transport efficiency of the GSF instability in stars. The Reynolds stress is of order $\langle\tilde{u}_x \tilde{u}_y \rangle H^{-1}$, where $H\mathrm{\equiv} U_0 {(\partial_x U_0)}^{-1}$ is the scale height of the mean flow $U_0 \mathrm{=} -\mathcal{S} x$. The time scale for modifying the flow is $\tau_\mathrm{turb}\sim U_0 H/ \langle\tilde{u}_x \tilde{u}_y \rangle\sim S x^2 
 \Omega^{-1} d^{-2} / \langle\tilde{u}_x \tilde{u}_y \rangle_\mathrm{dimensionless}$ \citep[see also][]{barker2019}. 
 
 Though $\mathrm{Pr}\sim10^{-6}$, the typical values of $r$ in the solar tachocline and red giant stars are  $r\sim 10^{-3}\textrm{--}1$ (varying with radius). Then, using Fig.~\ref{fig:f5} where $\langle\tilde{u}_x \tilde{u}_y \rangle_\mathrm{dimensionless}$ is on average $0.5$, we predict $\tau_\mathrm{turb} \Omega \sim 2S (x/d)^2$. This turbulent transport time scale is sufficiently short to be astrophysically important, depending on the relative length scale $x/d$ of the mean flow and shear strength $S$. For example, it can be as short as $\mathcal{O}(10)$ Myr using values of $S$ and $x/d$ for the solar tachocline. 
 
The turbulent transport rate depends sensitively also on the shear parameter $S$ (and latitude and orientation of the shear, i.e., radial or mixed radial-horizontal shear), and orders of magnitude faster turbulent transport is possible. We highlight that, because the turbulent time scale for the GSF instability can be shorter than $\mathcal{O}(10)$ Myr, incorporation of our transport model for the GSF turbulence in stellar evolution codes is warranted (particularly if extended to non-equatorial and 3-D GSF instabilities). Using such, the long-term impact on the evolution of the rotation profile may be assessed, informing us of the effects of the GSF instability in rapidly rotating young stars. In this regard, the transport model built here for the 2.5D equatorial GSF instability (and the thermohaline instability) is significant, as a reliable and reduced numerical treatment of the GSF instability-driven turbulence is now available.

\subsection{Relation to the thermohaline instability}
We emphasize that the equations describing 2.5-D thermohaline and GSF instabilites at the equator are identical, when the compositional diffusivity is equal to the kinematic viscosity---a case often realized in stars. We first write Eqs.~(7)--(9) of \cite{brown2013}, where they measure distance in units of the characteristic length scales $d$ of the fingers, time in units of the characteristic diffusion time scale $\tau\mathrm{=}d^2/\kappa$, and scaled temperature $T$ in units of $N^2 d$ [we label their $x$-coordinate with our $z$-coordinate and vice-versa]
\begin{subequations}
    \begin{align} \label{eq:brownuxeqn}
        \mathrm{Pr}^{-1} D u_x^\mathrm{B} &= -\partial_x p^\mathrm{B}  + (T^\mathrm{B}-\mu^\mathrm{B}) + \nabla^2 u_x^\mathrm{B},\\
        \mathrm{Pr}^{-1} D u_z^\mathrm{B} &= -\partial_z p^\mathrm{B}  + \nabla^2 u_z^\mathrm{B},\\
        D T^\mathrm{B} &= -u_x^\mathrm{B}  + \nabla^2 T^\mathrm{B},\\
        D \mu^\mathrm{B} &= -\frac{u_x^\mathrm{B}}{R_0} + \Pr \nabla^2 \mu^\mathrm{B},  \label{eq:brownmeqn}
    \end{align}
\end{subequations}
where the state vector $[u_x^\mathrm{B}, u_z^\mathrm{B}, T^\mathrm{B}, \mu^\mathrm{B}, p^\mathrm{B}]$ represents Brown-normalized $x$- and $z$-velocities, scaled temperature, chemical concentration, and fluid pressure, respectively. The so-called density ratio $R_0\mathrm{=}-N^2/\kappa_\mathrm{ep}^2$ may be recast as $R_0  \mathrm{=} 1 + r(\mathrm{Pr}^{-1}-1)$. The solutions of Eqs.~\eqref{eq:brownuxeqn}--\eqref{eq:brownmeqn} critically depend only on two parameters: Pr and $r$. Equations~\eqref{eq:brownuxeqn}--\eqref{eq:brownmeqn} for the thermohaline instability with the state vector $[u_x^\mathrm{B}, u_z^\mathrm{B}, T^\mathrm{B}, \mu^\mathrm{B}]$ are identical to Eqs.~\eqref{eq:dtu} and \eqref{eq:dtth} for the GSF instability with the state vector $[u_x, u_z, \theta, u_y]$, when we note 
\begin{subequations}
\begin{align}
u_x^\mathrm{B} &= \frac{u_x}{d/\tau},\\
u_z^\mathrm{B} &= \frac{u_z}{d/\tau},\\ 
T^\mathrm{B} &= \frac{\theta}{N^2 d},\\
\mu^\mathrm{B} &= -\frac{2 \tau \Omega }{\mathrm{Pr}} \frac{u_y}{d/\tau}.
\end{align}
\end{subequations}
Hence, the transport rates with two kinds of non-dimensionalizations---one using $\tau$ and $d$ as the characteristic time scale and length scale,  and another using $\Omega$ and $d$ as the relevant scales---are related in the manner
\begin{subequations}
\begin{align}\label{eq:figcaptionm}
    -\langle \widetilde{u}_x^{\mathrm{B}} \widetilde{\mu}^{\mathrm{B}} \rangle = \langle \widetilde{u}_x \widetilde{u}_y \rangle  \times 2\Omega \mathrm{Pr}^{1/2} (N^2)^{{-3/2}} d^{-2}, \\ \label{eq:figcaptionT}
    \langle \widetilde{u}_x^{\mathrm{B}} \widetilde{T}^{\mathrm{B}} \rangle = \langle \widetilde{u}_x \widetilde{\theta} \rangle  \times \mathrm{Pr}^{1/2} (N^2)^{{-3/2}} d^{-2},
\end{align}
\end{subequations}
where the variables with the superscripted `B' are functions of two essential parameters --- $r$ and Pr only, whereas the variables without the superscript are functions defined by the parameters $\mathrm{Pr}, S,$ and $N^2$.

The expressions given in Eqs.~\eqref{eq:figcaptionm} and \eqref{eq:figcaptionT} are plotted in Fig.~\ref{fig:f4}.  Thus our Fig.~\ref{fig:f4} also represents a comparison of chemical and heat transport between direct numerical simulations and analytical models for the thermohaline instability-driven turbulence.

The reduction of the governing equations of the GSF instability to two parameters ($r,\mathrm{Pr}$) is realized only in 2.5-D equatorial case, which is where the analogy of the GSF instability with the thermohaline instability becomes exact.

\section{Discussion and conclusions}\label{sec:discuss}
Turbulent transport in stellar interiors is a phenomenon too complex to represent directly in stellar evolution models. It is often parametrized using low-order models, 
such as mixing-length theories, or models that predict transport rates based on the fastest-growing unstable mode \citep[e.g.,][]{D2010,brown2013,barker2019}. The reliability of such models can be compromised by several factors: first, the instability dispersion relation is often anisotropic, a property that affects the nonlinear energy transfer, thereby driving low-frequency fluctuations (Fig.~\ref{fig:f2}). Second, nonlinear mode coupling can strongly excite more weakly growing unstable modes over a wide range of wavenumbers, presenting difficulties to single-mode theory-based predictions. To circumvent such challenges, we build a nonlinear mode-coupling theory, informed by detailed analyses of direct numerical simulations, to arrive at a reliable and analytic transport model, that is free of tunable parameters. We achieve this here for axisymmetric low-Pr turbulence driven by centrifugally unstable differential rotation (the GSF instability) at the equator in a stellar radiative zone.

Although 2.5-D turbulence driven by the GSF instability can differ from fully 3-D cases \citep[][]{barker2019}, strong secondary flows or jets have been found in 3-D global systems, also.  For example, recent simulations of fully 3-D spherical-shell non-rotating fingering convection have exhibited strong, large-scale jets \citep{tassin2023}. Such coherent jets are ubiquitious in various settings such as in geo- and astrophysical observations, numerical simulations of 3-D shear-flow instability-driven turbulence \citep{tripathi2023b}, and in laboratory fusion plasmas \citep{terry2019}. The examples show the large-scale flows can emerge even in global geometry. Hence, the success of our theory offers a future possibility to extend the statistical closure framework, presented here, to the more realistic 3-D simulations of the GSF instability, ideally in a sphere, at a general latitude and for a range of Pr \citep{barker2020,dymott2023, GB2015}. It is also possible that our closure-model framework can be adapted to magnetized turbulence driven by unstable differential rotation.

Since our formulae are fully analytic, they are quick to implement in stellar evolution codes such as \texttt{MESA} \citep{MESA1,MESA5} to reliably predict axisymmetric GSF instability-driven turbulent transport rates that vary with Pr and $r$ at different spatial grid points in an evolving star. It is straightforward to compute the values of $r$ and Pr at a given spatial grid point in a modeled star, and simply look up transport rates using our Fig.~\ref{fig:f5}(a) to prescribe the rates of transport of angular momentum and heat; to access the look up table of transport, see Data Availability. Since we have also provided formulae and frameworks for the turbulent transport of the axisymmetric GSF-analogous thermohaline instability, there now exists a reliable chemical transport model, which employs DNS-confirmed key elements of nonlinear saturation of the instability in stars.

\section*{Acknowledgments}
We thank the anonymous reviewer for constructive feedback. This work was enabled by the support from the US Department of Energy Grant No.~DE-SC0022257, from the STFC Grant Nos.~ST/S000275/1 and ST/W000873/1, and from the NASA HTMS Grant No.~80NSSC20K1280. A.E.F. acknowledges support from the George Ellery Hale Postdoctoral Fellowship in Solar, Stellar and Space Physics at the University of Colorado. Computations were performed using ARC4, part of the High Performance Computing facilities at the University of Leeds, and the DiRAC Data Intensive service at Leicester, operated by the University of Leicester IT Services, which forms part of the STFC DiRAC HPC Facility (\href{www.dirac.ac.uk}{www.dirac.ac.uk}). The equipment was funded by BEIS capital funding via STFC capital grants ST/K000373/1 and ST/R002363/1 and STFC DiRAC Operations grant ST/R001014/1. DiRAC is part of the National e-Infrastructure. We also used ACCESS computing resources via Allocation No.~TG-PHY130027. The inception of this work took place at the stimulating KITP program ``Probes of Transport in Stars” (supported in part by the NSF under Grant No.~NSF PHY-1748958).

\section*{Data Availability}
The GitHub repository, \\
\href{https://github.com/BindeshTripathi/GSF_transport}{https://github.com/BindeshTripathi/GSF\_transport}, hosts the data presented in Fig.~\ref{fig:f5}(a), and a python script where the closure model is implemented. Other data used in this article will be shared on reasonable request to the corresponding authors. 

\clearpage

\onecolumngrid

\setcounter{equation}{0}
\setcounter{figure}{0}
\setcounter{table}{0}
\makeatletter
\renewcommand{\theequation}{A\arabic{equation}}
\renewcommand{\thefigure}{A\arabic{figure}}

\section*{Appendix~A: Nonlinear mode-coupling coefficients of the GSF instability}\label{sec:appendixA}

The $j^\mathrm{th}$ eigenvector of the linear operator $\mathbf{L}$ of the GSF instability satisfies
\begin{equation}
    \hat{u}_{z,j} = -\frac{k_x}{k_z} \hat{u}_{x,j};\hspace{1cm} \hat{\theta}_j = -\frac{N^2}{\gamma_{\kappa,j}} \hat{u}_{x,j};\hspace{1cm} \hat{u}_{y,j} = -\frac{2\Omega-S}{\gamma_{\nu,j}} \hat{u}_{x,j},
\end{equation}
using which the matrix $\mathbf{E}$ of eigenvectors can be created and inverted as mentioned in the penultimate paragraph of Sec.~\ref{sec:eigmode}. To expedite analytic calculations, one may solve for the adjoint solutions $\mathbf{Y}_j$ of $\mathbf{L}$, which are the eigenvectors of $\mathbf{L}^\dagger$.  Such adjoint solutions $\mathbf{Y}_j$ form a biorthogonal basis with the eigenvectors $\mathbf{X}_j$ of $\mathbf{L}$ \citep{fraser2021, tripathi2022a, tripathi2023a, tripathi2023b}. That is, $\langle \mathbf{Y}_m, \mathbf{X}_j\rangle = {\mathbf{Y}_m^{\mathrm{T}\ast}} \mathbf{X}_j \propto \delta_{j,m}$, where $\mathrm{T}\ast$ is the transpose-conjugation operation; see Appendix~A of \cite{tripathi2022b} for a general mathematical proof.  
It is known that both $\mathbf{L}$ and $\mathbf{L}^\dagger$ have the same eigenvalues. The expedient adjoint technique is equivalent to inverting the matrix $\mathbf{E}$ of size $4\times 4$, and they deliver completely identical results, which we have verified. The $j^\mathrm{th}$ adjoint solution satisfies
\begin{equation}
    \hat{u}_{z,j} = 0;\hspace{1cm} \hat{\theta}_j = \frac{k_z^2}{k^2\gamma_{\kappa,j}} \hat{u}_{x,j};\hspace{1cm} \hat{u}_{y,j} = \frac{2\Omega k_z^2}{k^2\gamma_{\nu,j}} \hat{u}_{x,j}.
\end{equation}
Using such, we write the nonlinear mode-coupling coefficient between the two eigenmodes $m$ at wavenumber $\mathbf{k}'$ and $n$ at $\mathbf{k}''$, impacting the eigenmode $j$ at $\mathbf{k}$, as
\begin{equation}
    C_{jmn}^{(\mathbf{k},\mathbf{k}')} = \frac{\langle \mathbf{Y}_j(\mathbf{k}), \mathbf{N}(\mathbf{X}_m', \mathbf{X}_n'')\rangle}{\langle \mathbf{Y}_j(\mathbf{k}), \mathbf{X}_j(\mathbf{k})},
\end{equation}
where $\mathbf{N}(\mathbf{X}_m', \mathbf{X}_n'')$ is the nonlinearity vector; for example, its $\theta$-component is $-\mathbf{u}_m'\cdot \nabla \theta_n''$.

Following this procedure, we have distilled the analytic coupling coefficients for the GSF instability, and provide below their final expressions:
\begin{equation}\label{eq:Z11GSF}
        \dvec C_{\Z11}^{(\mathbf{k}, \mathbf{k}')}  = \frac{i \left(k_x'^2 -k_x''^2\right)}{k_z'f'f''},
\end{equation}
where $f$ is the eigenmode normalization factor. With inverse dimensions of $\hat{u}_{x,j}$, a suitable mode normalization can be $f' \mathrm{=}k' /\gamma'$ or $f'\mathrm{=}k'^2 d /(\gamma'  \gamma_\nu' \tau)$ or their variants, with $\tau\mathrm{=}d^2/\kappa$ as the characteristic diffusion time scale. Equation~\eqref{eq:Z11GSF} is simple as it represents the coupling between two unstable modes that drive the latitudinal flow $\Z$. The second coupling coefficient required in the closure model is
\begin{equation}\label{eq:1Z1GSF}
        \dvec C_{1\Z1}^{(\mathbf{k}'', \mathbf{k})}  = i k_z' \frac{\left[\frac{k''^2}{N^2 k_z''^2}\left(1-\frac{2k_x k_x''}{k''^2}\right) + \frac{1}{R_0 \gamma_\nu' \gamma_\nu''} - \frac{1}{\gamma_\kappa' \gamma_\kappa''} \right] f'' }{\left[\frac{k''^2}{N^2 k_z''^2} + \frac{1}{R_0 \gamma_\nu''^2} - \frac{1}{\gamma_\kappa''^2} \right] f'}.
\end{equation}
We note that the coupling coefficients depend only on wavenumbers and the input parameters such as 
$N^2$, Pr, $R_0\mathrm{=}-N^2/\kappa_\mathrm{ep}^2$ and the growth rates; the growth rates in turn depend only on wavenumbers and the input parameters [Eq.~\eqref{eq:growthrateeqn}]. Asymptotic approximation to the growth rate in the limit of, e.g., small Pr is possible \citep{brown2013}.

\section*{Appendix~B: Nonlinear mode-coupling coefficients of the Thermohaline instability}\label{sec:appendixB}

Here we provide analytic expressions needed for the closure model applicable to the thermohaline instability. 

We find the $j^\mathrm{th}$ eigenvector of the linear operator $\mathbf{L}$ found from Eqs.~\eqref{eq:brownuxeqn}--\eqref{eq:brownmeqn} for the thermohaline instability satisfies
\begin{equation}
    \hat{u}^\mathrm{B}_{z,j} = -\frac{k_x}{k_z} \hat{u}^\mathrm{B}_{x,j};\hspace{1cm} \hat{T}^\mathrm{B}_j = -\frac{1}{\gamma_{\kappa,j}} \hat{u}^\mathrm{B}_{x,j};\hspace{1cm} \hat{\mu}^\mathrm{B}_j = -\frac{1}{R_0 \gamma_{\nu,j}} \hat{u}^\mathrm{B}_{x,j}.
\end{equation}
The $j^\mathrm{th}$ adjoint solution for the thermohaline instability is
\begin{equation}
    \hat{u}^\mathrm{B}_{z,j} = 0;\hspace{1cm} \hat{T}^\mathrm{B}_j = \frac{k_z^2 \mathrm{Pr} }{k^2\gamma_{\kappa,j}} \hat{u}^\mathrm{B}_{x,j};\hspace{1cm} \hat{\mu}^\mathrm{B}_j = -\frac{k_z^2 \mathrm{Pr}}{k^2\gamma_{\nu,j}} \hat{u}^\mathrm{B}_{x,j}.
\end{equation}

The coupling coefficients for the thermohaline instability are identically the same as those for the GSF instability; Equations~\eqref{eq:Z11GSF} and \eqref{eq:1Z1GSF} require only a minute modification: $N^2$ appearing twice in Eq.~\eqref{eq:1Z1GSF} should be replaced with Pr. We, conclude that our closure model and predictions are directly applicable to the thermohaline instability, as well.  This is significant because the turbulent transport efficiencies of these two instabilites in stars are not known, but are generally thought to be important. With the closure model at hand, we can now make predictions reliably for both instabilities.\\

\section*{Appendix~C: Details of Closure Model Calculations}\label{sec:appendixC}

To make the statistical closure model more accessible to a wide range of readers, we provide below detailed, step-by-step derivations.

\subsection*{C1.~Amplitude evolution equation}
The mode-amplitude $\beta_j$ evolution equation, given in Eq.~\eqref{eq:betaevoln}, for a wavenumber $\mathbf{k}= (k_x,k_z\mathrm{\neq} 0)$, with $j=1, 2,$ or $3$, is
\begin{equation}\label{eq:b10beta}
\begin{aligned}[b]
    \partial_t \beta_j = \gamma_j \beta_j + \sum_{\substack{\bm{k}',m,n}} C_{jmn}^{(\bm{k},\bm{k}')}  \beta_m'\beta_n'' +  \sum_{\bm{k}',n} \Bigg\{ 
    &\left[ C_{j \Y n}^{(\bm{k},\bm{k}')} + C_{j n \Y}^{(\bm{k},\bm{k}'')}  \right] \Y' \beta_n''\\
    &+\left[ C_{j \Z n}^{(\bm{k},\bm{k}')} + C_{j n \Z}^{(\bm{k},\bm{k}'')}  \right] \Z' \beta_n''\\
    &+\left[ C_{j \T n}^{(\bm{k},\bm{k}')} + C_{j n \T}^{(\bm{k},\bm{k}'')}  \right] \T' \beta_n''\Bigg\},
\end{aligned}
\end{equation}
whereas, at $\mathbf{k}= (k_x,k_z\mathrm{=} 0)$, one finds that the fluctuation amplitude evolves according to Eq.~\eqref{eq:Fevoln}, which is expanded below
\begin{subequations}
\begin{align} \label{eq:b10Y}
    \partial_t \Y &= -\gamma_\Y \Y + \sum_{\bm{k}',m,n: k_z\mathrm{=}0} C_{\Y m n}^{(\bm{k},\bm{k}')} \beta_m'\beta_n'',\\ \label{eq:b10Z}
    \partial_t \Z &= -\gamma_\Z \Z + \sum_{\bm{k}',m,n: k_z\mathrm{=}0} C_{\Z m n}^{(\bm{k},\bm{k}')} \beta_m'\beta_n'',\\ \label{eq:b10T}
    \partial_t \T &= -\gamma_\T \T + \sum_{\bm{k}',m,n: k_z\mathrm{=}0} C_{\T m n}^{(\bm{k},\bm{k}')} \beta_m'\beta_n'',
\end{align}
\end{subequations}
where $\gamma_\Y=\gamma_\Z=\nu k^2$ and $\gamma_\T = \kappa k^2$ are the damping rates.

\subsection*{C2.~Eigenmode-energy evolution}
To derive an evolution equation for energy in the $j^\mathrm{th}$ eigenmode at $\mathbf{k}= (k_x,k_z\mathrm{\neq} 0)$, we multiply Eq.~\eqref{eq:b10beta} with $\beta_j^\ast$ and add a complex conjugate of the resulting equation to arrive at
\begin{equation} 
\begin{aligned}[b] \label{eq:energyevoln1}
    \partial_t |\beta_j|^2 &= 2\Real \gamma_j |\beta_j|^2 +  \sum_{\substack{\bm{k}',m,n}} 2\Real \left[ C_{jmn}^{(\bm{k},\bm{k}')} \langle  \beta_m'\beta_n'' \beta_j^\ast \rangle \right] \\
    &\hspace{2.5cm}+ \sum_{\bm{k}'} 2\Real \Bigg\{ \left[ C_{j\Y j}^{(\bm{k},\bm{k}')} \langle \Y' \beta_j'' \beta_j^\ast \rangle + C_{j\Z j}^{(\bm{k},\bm{k}')} \langle \Z' \beta_j'' \beta_j^\ast \rangle + C_{j\T j}^{(\bm{k},\bm{k}')} \langle \T' \beta_j'' \beta_j^\ast \rangle  \right]\Big\vert_{k_z'\mathrm{=}0} \Bigg. \\
    &\hspace{4.34cm} \Bigg. + \left[  C_{jj\Y}^{(\bm{k},\bm{k}')} \langle \beta_j'  \Y'' \beta_j^\ast \rangle  + C_{jj\Z}^{(\bm{k},\bm{k}')} \langle \beta_j'  \Z''  \beta_j^\ast \rangle + C_{jj\T}^{(\bm{k},\bm{k}')} \langle \beta_j'  \T'' \beta_j^\ast \rangle  \right]\Big\vert_{k_z'\mathrm{=}k_z} \Bigg\}.
\end{aligned}
\end{equation}

Similar equations can be derived for fluctuation energy at GSF-stable wavenumbers $\mathbf{k}= (k_x,k_z\mathrm{=} 0)$ using Eqs.~\eqref{eq:b10Y}--\eqref{eq:b10T}:
\begin{subequations} 
\begin{align} \label{eq:energyevoln2}
    \partial_t |\Y|^2 &= -2\Real \gamma_\Y |\Y|^2 + \sum_{\bm{k}',m}2\Real \left[ C_{\Y mm}^{(\bm{k},\bm{k}')} \langle \beta_m'\beta_m'' \Y^\ast \rangle \right]\Big\vert_{k_z\mathrm{=}0},\\ \label{eq:energyevoln3}
    \partial_t |\Z|^2 &= -2\Real \gamma_\Z |\Z|^2 + \sum_{\bm{k}',m} 2\Real \left[C_{\Z mm}^{(\bm{k},\bm{k}')} \langle \beta_m'\beta_m'' \Z^\ast \rangle \right]\Big\vert_{k_z\mathrm{=}0},\\ \label{eq:energyevoln4}
    \partial_t |\T|^2 &= -2\Real \gamma_\T |\T|^2 + \sum_{\bm{k}',m} 2\Real \left[C_{\T mm}^{(\bm{k},\bm{k}')} \langle \beta_m'\beta_m'' \T^\ast \rangle \right]\Big\vert_{k_z\mathrm{=}0}.
\end{align}
\end{subequations}

Since numerical simulations inform us that the triplets with a latitudinal flow at $(k_x,0)$, i.e., $\Z$, dominates the nonlinear energy transfer, we may drop nonlinear terms on the right-hand side of Eq.~\eqref{eq:energyevoln1} that do not involve $\Z$. In the resulting equation, because $\Y$ and $\T$ do not appear, Eqs.~\eqref{eq:energyevoln2} and \eqref{eq:energyevoln4} can also be removed, which allows us to write the following set of equations:
\begin{subequations} 
\begin{align} \label{eq:energyevoln1new}
    \partial_t |\beta_j|^2 &= 2\Real \gamma_j |\beta_j|^2 + \sum_{\bm{k}'} 2\Real \Bigg\{ \left[ C_{j\Z j}^{(\bm{k},\bm{k}')} \langle \Z' \beta_j'' \beta_j^\ast \rangle   \right]\Big\vert_{k_z'\mathrm{=}0} + \left[   C_{jj\Z}^{(\bm{k},\bm{k}')} \langle \beta_j'  \Z''  \beta_j^\ast \rangle \right]\Big\vert_{k_z'\mathrm{=}k_z} \Bigg\},\\
    \label{eq:energyevoln3new}
    \partial_t |\Z|^2 &= -2\Real \gamma_\Z |\Z|^2 + \sum_{\bm{k}',m} 2\Real \left[C_{\Z mm}^{(\bm{k},\bm{k}')} \langle \beta_m'\beta_m'' \Z^\ast \rangle \right]\Big\vert_{k_z\mathrm{=}0}.
\end{align}
\end{subequations}

\subsection*{C3.~Triplet correlation evolution}

To obtain evolution equations for the terms on the right-hand side of Eqs.~\eqref{eq:energyevoln1new} and \eqref{eq:energyevoln3new}, we multiply Eq.~\eqref{eq:betaevoln} with two amplitudes.  For example, to determine the evolution of $\langle \Z' \beta_j'' \beta_j^\ast \rangle$, first, we multiply Eq.~\eqref{eq:b10beta} for $\beta_j$ with $\Z'\beta_j''$; second, we multiply another equation, similar to Eq.~\eqref{eq:b10beta}, but for $\beta_j''$ with $\Z'\beta_j^\ast$; and, finally, we multiply Eq.~\eqref{eq:b10Z}, for $\Z'$, with $\beta_j'' \beta_j^\ast$, and add them all together. Using this, we provide below an example evolution equation for triplet correlations:
\begin{equation}\label{eq:b11}
\begin{aligned}[b]
    \left[\partial_t - \left( -\gamma_{\Z}' +\gamma_j'' +\gamma_j^\ast \right) \right] \langle {\Z}' \beta_j''\beta_j^\ast \rangle\vert_{k_z'\mathrm{=}0} 
    =&
    \Bigg\{\sum_{\bm{k}''', m}\left[C_{{\Z} mm}^{(\bm{k}', \bm{k}''')} \langle \beta_m'''\beta_m(\bm{k}'-\bm{k}''')\beta_j''\beta_j^\ast \rangle\right]\\
    &+\sum_{k_x'''}\left[C_{j\Y j}^{(\bm{k}-\bm{k}', \bm{k}''')} \langle \Y''' \beta_j(\bm{k}-\bm{k}'-\bm{k}''') {\Z}' \beta_j^\ast \rangle\vert_{k_z'''\mathrm{=}0}\right.\\
    &\hspace{0.9cm}+ C_{j\Z j}^{(\bm{k}-\bm{k}', \bm{k}''')} \langle \Z''' \beta_j(\bm{k}-\bm{k}'-\bm{k}''') {\Z}' \beta_j^\ast \rangle\vert_{k_z'''\mathrm{=}0}\\
    &\hspace{0.9cm}+ C_{j\T j}^{(\bm{k}-\bm{k}', \bm{k}''')} \langle \T''' \beta_j(\bm{k}-\bm{k}'-\bm{k}''') {\Z}' \beta_j^\ast \rangle\vert_{k_z'''\mathrm{=}0}\\
    &\hspace{0.9cm}+C_{jj\Y}^{(\bm{k}-\bm{k}', \bm{k}''')} \langle \beta_j''' \Y(\bm{k}-\bm{k}'-\bm{k}''') {\Z}' \beta_j^\ast \rangle\vert_{k_z'''\mathrm{=}k_z}\\
    &\hspace{0.9cm}+ C_{jj\Z }^{(\bm{k}-\bm{k}', \bm{k}''')} \langle \beta_j''' \Z(\bm{k}-\bm{k}'-\bm{k}''') {\Z}' \beta_j^\ast \rangle\vert_{k_z'''\mathrm{=}k_z}\\
    &\hspace{0.9cm}\left.+ C_{jj\T }^{(\bm{k}-\bm{k}', \bm{k}''')} \langle \beta_j''' \T(\bm{k}-\bm{k}'-\bm{k}''') {\Z}' \beta_j^\ast \rangle\vert_{k_z'''\mathrm{=}k_z} \right]\\
    &+\sum_{k_x'''}\left[C_{j\Y j}^{(\bm{k}, \bm{k}''')\ast} \langle \Y'''^\ast \beta_j^\ast(\bm{k}-\bm{k}''') {\Z}' \beta_j'' \rangle\vert_{k_z'''\mathrm{=}0}\right.\\
    &\hspace{0.9cm}+ C_{j\Z j}^{(\bm{k}, \bm{k}''')\ast} \langle \Z'''^\ast \beta_j^\ast(\bm{k}-\bm{k}''') {\Z}' \beta_j'' \rangle\vert_{k_z'''\mathrm{=}0}\\
    &\hspace{0.9cm}+ C_{j\T j}^{(\bm{k}, \bm{k}''')\ast} \langle \T'''^\ast \beta_j^\ast(\bm{k}-\bm{k}''') {\Z}' \beta_j'' \rangle\vert_{k_z'''\mathrm{=}0}\\
    &\hspace{0.9cm}+C_{jj\Y}^{(\bm{k}, \bm{k}''')\ast} \langle \beta_j'''^\ast \Y^\ast(\bm{k}-\bm{k}''') {\Z}' \beta_j'' \rangle\vert_{k_z'''\mathrm{=}k_z}\\
    &\hspace{0.9cm}+ C_{jj\Z }^{(\bm{k}, \bm{k}''')\ast} \langle \beta_j'''^\ast \Z^\ast(\bm{k}-\bm{k}''') {\Z}' \beta_j'' \rangle\vert_{k_z'''\mathrm{=}k_z}\\
    &\hspace{0.9cm}\Bigg.\left.\left.+ C_{jj\T }^{(\bm{k}, \bm{k}''')\ast} \langle \beta_j'''^\ast \T^\ast(\bm{k}-\bm{k}''') {\Z}' \beta_j'' \rangle\vert_{k_z'''\mathrm{=}k_z} \right]
    \Bigg\}\right|_{k_z'\mathrm{=}0}.\\
\end{aligned}
\end{equation}

Since the numerical simulations show dominant coupling between the latitudinal flow $\Z$ and two unstable modes in a bath of turbulent interactions, the interactions involving unstable modes only can be excluded. This immediately implies that the first term on the right-hand side of Eq.~\eqref{eq:b11} can be dropped in the face of remaining dominant terms. Among the remaining terms, the fourth-order correlations that have different components of velocity, e.g., the terms where $\Y$ and $\Z$ appear together on the right-hand side of Eq.~\eqref{eq:b11}, do not form terms that appear in the definition of energy. Guided by numerical simulations where nonlinear coupling to $\Y$ and $\T$ are unimportant, only the energy(-like) terms with $\Z$ will be kept henceforth. The resulting equation then reads
\begin{equation}\label{eq:b12}
\begin{aligned}[b]
    \left[\partial_t - \left( -\gamma_\Z' +\gamma_j'' +\gamma_j^\ast \right) \right] \langle \Z' \beta_j''\beta_j^\ast \rangle\vert_{k_z'\mathrm{=}0} 
    =&
    \Bigg\{\sum_{k_x'''}\left[C_{j\Z j}^{(\bm{k}-\bm{k}', \bm{k}''')} \langle \Z''' \beta_j(\bm{k}-\bm{k}'-\bm{k}''') \Z' \beta_j^\ast \rangle\vert_{k_z'''\mathrm{=}0}\right.\\
    &\hspace{0.9cm}+C_{jj\Z}^{(\bm{k}-\bm{k}', \bm{k}''')} \langle \beta_j''' \Z(\bm{k}-\bm{k}'-\bm{k}''') \Z' \beta_j^\ast \rangle\vert_{k_z'''\mathrm{=}k_z}\\
    &\hspace{0.9cm}+C_{j\Z j}^{(\bm{k}, \bm{k}''')\ast} \langle \Z'''^\ast \beta_j^\ast(\bm{k}-\bm{k}''') \Z' \beta_j'' \rangle\vert_{k_z'''\mathrm{=}0}\\
    &\hspace{0.9cm}\left.\Bigg.\left.+C_{jj\Z}^{(\bm{k}, \bm{k}''')\ast} \langle \beta_j'''^\ast \Z^\ast(\bm{k}-\bm{k}''') \Z' \beta_j'' \rangle\vert_{k_z'''\mathrm{=}k_z} \right]
    \Bigg\}\right|_{k_z'\mathrm{=}0}.\\
\end{aligned}
\end{equation}

The same procedure is then repeated to find evolution equations for the other two triplet correlations that appear in 
Eqs.~\eqref{eq:energyevoln1new} and \eqref{eq:energyevoln3new}.

\subsection*{C4: Quadruplet correlations and Statistical closure approximation} 
Equation~\eqref{eq:b12} can be solved using the technique of Green's function inversion and Markovianization, a standard step in EDQNM closure [although, here, we do not modify the growth rate $\gamma$'s with the amplitude-dependent nonlinear frequency, an approximation justifiable for the low-wavenumber regime; for more details, see \citet{terry2018}]. Such a solution yields
\begin{equation} \label{eq:b13}
\begin{aligned}[b]
    \langle \Z' \beta_j''\beta_j^\ast \rangle\vert_{k_z'\mathrm{=}0} 
    = - \Bigg\{ \left( -\gamma_\Z' +\gamma_j'' +\gamma_j^\ast \right)^{-1} |\Z'|^2 & \Big[ \left( C_{j\Z j}^{(\bm{k}'', -\bm{k}')} +  C_{jj\Z }^{(\bm{k}'', \bm{k})}  \right) |\beta_j|^2 + \left( C_{j\Z j}^{(\bm{k},\bm{k}')\ast} +  C_{jj\Z }^{(\bm{k}, \bm{k}'')\ast}  \right) |\beta_j''|^2 \Big] \Bigg\}\Bigg\vert_{k_z'\mathrm{=}0}.
\end{aligned}
\end{equation}


Similarly, the other two triplet correlations that appear in Eqs.~\eqref{eq:energyevoln1new} and \eqref{eq:energyevoln3new} can also be solved to obtain
\begin{equation}\label{eq:b14}
\begin{aligned}[b]
    \langle \beta_j' \Z'' \beta_j^\ast \rangle\vert_{k_z'\mathrm{=}k_z} 
    = - \Bigg\{ \left( \gamma_j' -\gamma_\Z'' +\gamma_j^\ast \right)^{-1} |\Z''|^2 &\Big[ \left( C_{j\Z j}^{(\bm{k}', -\bm{k}'')} +  C_{jj\Z }^{(\bm{k}', \bm{k})}  \right) |\beta_j|^2 + \left( C_{j\Z j}^{(\bm{k},\bm{k}'')\ast} +  C_{jj\Z }^{(\bm{k}, \bm{k}')\ast}  \right) |\beta_j'|^2 \Big] \Bigg\}\Bigg\vert_{k_z'\mathrm{=}k_z},
\end{aligned}
\end{equation}
and
\begin{equation}\label{eq:b15}
\begin{aligned}[b]
    \langle \beta_j' \beta_j'' \Z^\ast \rangle\vert_{k_z\mathrm{=}0} 
    = - \Bigg\{ \left( \gamma_j' +\gamma_j'' -\gamma_\Z^\ast \right)^{-1} |\Z|^2 &\Big[ \left( C_{j\Z j}^{(\bm{k}', \bm{k})} +  C_{jj\Z }^{(\bm{k}', -\bm{k}'')}  \right) |\beta_j''|^2 + \left( C_{j\Z j}^{(\bm{k}'',\bm{k})} +  C_{jj\Z }^{(\bm{k}'', -\bm{k}')}  \right) |\beta_j'|^2 \Big] \Bigg\}\Bigg\vert_{k_z\mathrm{=}0}.
\end{aligned}
\end{equation}

\subsection*{C5: A set of EDQNM-closed energy evolution equations}
Solutions of the triplet correlations from Eqs.~\eqref{eq:b13}--\eqref{eq:b15} are now substituted into Eqs.~\eqref{eq:energyevoln1new} and \eqref{eq:energyevoln3new}, which results in the following set of closed equations:
\begin{equation} 
\begin{aligned}[b] \label{eq:finalenergyevoln1}
    \hspace{-0.94cm}\partial_t |\beta_j|^2 = &2\Real \gamma_j |\beta_j|^2 - \sum_{\bm{k}'} 2\Real \Bigg[ \frac{ C_{j\Z j}^{(\bm{k},\bm{k}')} |\Z'|^2}{\left( -\gamma_\Z' +\gamma_j'' +\gamma_j^\ast \right)}  \left\{ \dvec C_{j\Z j}^{(\bm{k}'', -\bm{k}')} |\beta_j|^2 + \dvec C_{j\Z j}^{(\bm{k},\bm{k}')\ast} |\beta_j''|^2 \right\} \Bigg] \Bigg\vert_{k_z'\mathrm{=}0} \\
    &\hspace{1.95cm}-  \sum_{\bm{k}'} 2\Real \Bigg[  \frac{ C_{jj\Z}^{(\bm{k},\bm{k}')} |\Z''|^2 }{\left( \gamma_j' -\gamma_\Z'' +\gamma_j^\ast \right)} \left\{ \dvec C_{j\Z j}^{(\bm{k}', -\bm{k}'')} |\beta_j|^2 + \dvec C_{j\Z j}^{(\bm{k},\bm{k}'')\ast} |\beta_j'|^2 \right\}\Bigg]\Bigg\vert_{k_z'\mathrm{=}k_z},
\end{aligned}
\end{equation}

\begin{equation}
\label{eq:finalenergyevoln3}
    \partial_t |\Z|^2 =  -2\Real \gamma_\Z |\Z|^2 - |\Z|^2  \sum_{\bm{k}'}\sum_{j=1}^{3} 2\Real \Bigg. \left[\frac{ C_{\Z jj}^{(\bm{k},\bm{k}')} }{\left( \gamma_j' +\gamma_j'' -\gamma_\Z^\ast \right)} \left\{ \dvec C_{j\Z j}^{(\bm{k}', \bm{k})} |\beta_j''|^2 + \dvec C_{j\Z j}^{(\bm{k}'',\bm{k})} |\beta_j'|^2 \right\} \right]\Bigg\vert_{k_z\mathrm{=}0},
\end{equation}
where $\dvec C_{lmn}^{(\bm{p}, \bm{q})} = C_{lmn}^{(\bm{p},\bm{q})} + C_{lnm}^{(\bm{p},\bm{p}-\bm{q})}$ is the symmetrized coupling coefficient.

Equation~\eqref{eq:finalenergyevoln3} can be simplified by considering that it is the pairs of unstable modes ($j=1$) that excite the latitudinal flow $\Z$:
\begin{equation}\label{eq:dtz2appendix}
\partial_t |\Z|^2/2 
=  -\gamma_\Z |\Z|^2 + |\Z|^2  \sum_{\bm{k}'}|\beta_1'|^2 \,   \Real \Bigg. \left[ \frac{-\dvec C_{\Z 11}^{(\bm{k},\bm{k}')} \dvec C_{1\Z 1}^{(\bm{k}'',\bm{k})}}{\gamma_1' +\gamma_1'' -\gamma_\Z^\ast}  \right],
\end{equation}
where the term inside the wavenumber-summation in Eq.~\eqref{eq:finalenergyevoln3} has been symmetrized.






\bibliography{references_aastex}{}
\bibliographystyle{aasjournal}

\end{document}